\documentclass[
preprint,
jcp,
aip,
amsmath,amssymb,
]{revtex4-1}
\usepackage{placeins}
\usepackage{bm}
\usepackage{graphicx}
\usepackage{mathtools}
\usepackage{gensymb}
\usepackage[table]{xcolor}
\usepackage{array}
\usepackage{makecell}
\usepackage{sistyle}
\usepackage{amsfonts}
\usepackage{multirow}
\usepackage[utf8]{inputenc}
\usepackage[T1]{fontenc}
\usepackage{lmodern}
\usepackage[outdir=./,update]{epstopdf}
\usepackage[english]{babel}
\usepackage{graphicx}
\usepackage{caption}
\usepackage{subcaption}
\usepackage[margin=1.2 in]{geometry}

\begin{document}

\title{Reactivity of hydrated hydroxide anion clusters OH(H$_{2}$O)$_{n}^{-}$ with H and Rb: an \textit{ab initio} study.}

\author{Milaim Kas}
\email{milakas@ulb.ac.be}
\author{J\'er\^ome Loreau}
\author{Jacques Li\'evin}
\author{Nathalie Vaeck}
\affiliation{
 Service de Chimie Quantique et Photophysique (CQP)\\
 Universit\'e libre de Bruxelles (ULB), Brussels, Belgium 
}

\date{\today}

\begin{abstract}
We present a theoretical investigation of the hydrated hydroxide anion clusters OH(H$_{2}$O)$_{n}^{-}$ and of the collisional complexes H-OH(H$_{2}$O)$_{n}^{-}$ and Rb-OH(H$_{2}$O)$_{n}^{-}$ (with n$=1-4$). The MP2 and CCSD(T) methods are used to calculate interaction energies, optimized geometries and vertical detachment energies. Part of the potential energy surfaces are explored with a focus on the autodetachment region. We point out the importance of diffuse functions to correctly describe the latter. We use our results to discuss the different water loss and electronic detachment channels which are the main reaction routes at room temperature. In particular, we have considered a direct and an indirect process for the electronic detachment, depending on whether water loss follows or precedes the detachment of the excess electron. We use our results to discuss the implication for astrochemistry and hybrid trap experiments in the context of cold chemistry.     
\end{abstract}

\maketitle
\makeatletter
\let\toc@pre\relax
\let\toc@post\relax
\makeatother

\section{Introduction}
Molecular clusters are often seen as bridges between gas phase elementary reactions and condensed phase \cite{Castleman1996}. This is especially the case for water clusters. Numerous theoretical and experimental studies have been dedicated to the understanding of the structure and dynamics of neutral water clusters \cite{Erkoc2000}, hydrated ions \cite{Lalitha2014}, hydrated electrons \cite{Novakovskaya2002,Hao2003,Tachikawa2006} and protons \cite{Greve2010}, and anionic water clusters \cite{Neumark2008}. \\
The hydrated hydroxide anion clusters OH(H$_{2}$O)$_{n}^{-}$ are of particular importance in the field of solution chemistry, and they have been extensively studied by both theoretical \cite{Lee2004,Masamura2000,Lin2015,Egan2018} and experimental approaches \cite{Svendsen2004a,Robertson2003}. Various properties such as dissociation energies, ionization potential, harmonic frequencies, optimized structures, solvation energies and proton affinities have been obtained. 
In the present case, we focus on two different topics where the hydrated hydroxide anion clusters have only been sparsely studied : i) co-trapping (or hybrid trap) experiments in the context of cold chemistry and ii) astrochemistry. \\

\noindent i) The ability to confine and cool down atoms and molecules at temperatures below $\mu$K temperatures has lead to many interesting new area of research, ranging from precision spectroscopy \cite{Loh2013} and test of fundamental constants \cite{Bethlem2009} to quantum control of chemical reactions \cite{Quemener2012}. One of the possible method to cool down ionic species is through sympathetic cooling, where the ion is trapped and brought into contact with a cold buffer gas. The latter can be either a cold cryogenic atomic gas, such as He, or a laser cooled atomic cloud such as Rb. Even though the translational motion has been shown to effectively cool down, the decrease in the temperature of the internal motion (vibration, rotation) is more difficult to probe. Indeed, several factors such as micromotion \cite{Cetina2012,Holtkemeier2016}, reactive collisions with the buffer gas, collisions with background gas \cite{Ravi2012} and black body radiation prevent the ions from thermalizing with the buffer gas. Therefore, measurements of the internal state temperature may bring insights into the cooling dynamics. The OH(H$_{2}$O)$_{n}^{-}$ species present interesting features since they undergo both vibrational and rotational motions, can be easily produced and exhibit large vertical detachment energies \cite{Morita2014}. In addition, threshold photodetachment techniques can be used to probe their internal structure \cite{Otto2013}. Reactive collisions are also expected (when using  other species than rare gas atoms) and can be studied trough loss measurements \cite{Deiglmayr2012}. When dealing with anions, detachment reactions are usually one of the main reactive channels. Furthermore, comparison with OH$^{-}$ may help understand the effect of hydration on gas phase reactions. So far, molecular anions have only been sparsely studied in cold environment, however they are attracting increasing attention. \\    
           
\noindent ii) Several anions have been detected in various astrophysical environments: coma of comets \cite{Chaizy1991}, interstellar molecular clouds \cite{McCarthy2006}, and extraterrestrial atmospheres \cite{Vuitton2009}. In particular, since the OH$^{-}$ anion has been detected in the coma of comets \cite{Cordiner2014} and water molecules are predicted to form cluster in such environments \cite{Bykov2013, Crifo1990}, one could also anticipate the formation of hydrated hydroxide anion clusters. This species will certainly have an effect on chemical models which seem to underestimate the abundance of OH$^{-}$ \cite{Cordiner2014}. In addition, even tough molecular anions are not expected to exhibits long lifetimes, they may play important role in the formation of new neutral species. One of the main destruction pathway for anions is through associative electronic detachment reaction. Atomic hydrogen being the most abundant species in space, its reactivity is of particular importance.  \\      

\noindent The collision between hydroxide water clusters anions and an atomic specie M can lead to several outcomes: bond breaking, bond forming and/or electronic detachment. Since the O-H bond within the water cluster is rather strong, we have only considered water loss scenarios and M-OH bond forming along with electronic detachment process. The mechanism of this last process is not straightforward with two different pathways that should be considered: 
\begin{enumerate}
\item[(1)] M+OH(H$_{2}$O)$_{n}^{-} \rightarrow$ M-OH(H$_{2}$O)$_{n} + e^{-} \rightarrow $ MOH(H$_{2}$O)$_{l}$+$(n-l)\,$H$_{2}$O + $e^{-}$
\item[(2)] M+OH(H$_{2}$O)$_{n}^{-} \rightarrow$ M-OH(H$_{2}$O)$_{n}^{-} \rightarrow $ \\ 
MOH(H$_{2}$O)$_{l}^{-}+(n-l)\,$H$_{2}$O $\rightarrow $ MOH(H$_{2}$O)$_{k}+(n-k)\,$H$_{2}$O + $e^{-}$
\end{enumerate}  
Case $(1)$ can be seen as a direct associative electronic detachment reaction. If the autodetachment region is accessible at the considered collision energy, it is reached during the formation of the M-OH bond, leading to the ejection of the electron. The vibrational excess energy is then distributed among the H-bonds which may ultimately break, leading to water loss. Case $(2)$ corresponds to an indirect mechanism. If the autodetachment region is not accessible at the considered energy range, the energy of the formed M-OH bond will first lead to H-bond breaking. The ejection of the electron may then occur along the H-bond dissociation path. In particular, it may lead to the production of MOH(H$_{2}$O)$^{-}$ anions if their vertical detachment energy (VDE) is sufficiently large to bind the excess electron. This will strongly depend on the M atom. \\

We have investigated this reactive process for M=H (section \ref{sec_H}) and M=Rb (section \ref{sec_Rb}). \\

\noindent Experimental work on the reaction between H and OH(H$_{2}$O)$_{0-2-3}^{-}$ \cite{Howard1975} at 296 K have shown that the dominant channel is the associative electronic detachment (AED) with water loss reaction. All reactions are exothermic, which is due to the formation of the strong H-O bond of water during the associative detachment process. The energy released (about 5.1 eV \cite{Maksyutenko2006}) is enough to eject the electron and break all H-bond (ranging from 0.6 to 1 eV \cite{Masamura2000}) which leads to water loss. The obtained rate constant for the non hydrated case (H+OH$^{-}$) is very close to the Langevin rate while the rate decreases for increasing hydration number \cite{Howard1975}. This has been attributed to the steric effect of the water shell which prevents the H atom from approaching the OH group. Since no anions have been detected as products of the reaction, we can thus expect the electronic detachment process to be very efficient. However, as discussed above, the reaction can have either a direct or an indirect nature. Both cases will be investigated. AED reactions can either occur through "curve crossing" mechanisms where the neutral potential energy surface (PES) crosses the anion's or through non-adiabatic effects. In the first case, the rate of detachment is usually in the picosecond regime ($10^-13$s$-10^-15$s \cite{Simons2005}) while the rate of detachment is much smaller for the second case. Here we will mainly focus our discussion on the former mechanism. \\

\noindent The paper is structured as follows: in section \ref{sec_cluster} we investigate the molecular structure and VDE of the most stable hydrated hydroxide anions. We discuss their rotational and vibrational motion and make some prediction in the context of hybrid trap experiments. In section \ref{sec_H} we investigate the reactivity of H with the hydrated hydroxide anions: interaction potential and autodetachment region. We first focus on the smallest cluster H+OH(H$_{2}$O)$^{-}$ (section \ref{sec_Hn1}) before considering the case of larger cluster (section \ref{sec_Hn}). We then switch to the case of Rb. Finally, comparison are made between H and Rb where we have stressed on the context of astrochemistry and hybrid trap experiments, respectively.       
  
  \FloatBarrier
  
\section{Methods of calculation}

All calculations have been performed in $C_{1}$ symmetry (no element of symmetry) using the {\small MOLPRO} program \cite{MOLPRO}. The augmented correlation-consistent valence ntuple zeta basis sets aug-cc-pVnZ (shortened AVnZ) have been used for the O and H atoms. The $1s$ core electron of O atoms have been kept frozen (uncorrelated) in all calculations. For the Rb atom, the 28 core electrons are described by the ECP28MDF effective core potential \cite{} whereas the 4s and 4p outer core electrons and 5s valence electrons are described by its companion $spdfg$ basis set \cite{Lim2005}. Since the correlation of the outer core electrons has been shown to be important for alkali and alkali-earth atoms \cite{Kas2017}, we have included the 4s and 4p electrons of Rb into the correlation treatment. Second order Moller-Plesset perturbation theory (MP2) and coupled-cluster with single, double and perturbative tripe excitations (CCSD(T)) methods have been used and compared throughout the work. The overall level of theory will be denoted as method/O and H basis set/collisional H or Rb basis set. The effect of the basis set superposition error (BSSE) as well as the influence of additional diffuse functions have also been investigated. The BSSE is taken into account using the usual counterpoise approach \cite{Boys1970} and will be denoted by "+BSSE". The "+aug" notation will be used to denote extend basis set where $5s3p1d$ and $5s2d1f1g$ even tempered diffuse functions have been added to the basis set describing the collisional H and Rb atom, respectively.      

\section{OH(H$_{2}$O)$_{n}^{-}$ water cluster}
\label{sec_cluster}

We have investigated the most stable clusters up to 4 water molecules ($n=4$) by performing our geometry optimization at the MP2/AVTZ level of theory, starting from the optimized structures taken from \cite{Lee2004} where the same level of theory has been used. There is only one stable conformer for $n=1$ and $n=2$ but two for $n=3$. According to the literature \cite{Lee2004,Masamura2002a}, there are 3 to 5 isomers for $n=4$. We have only considered the 3 most stable ones. In addition we have have performed harmonic frequency calculations to ensure than our optimized structures correspond to a true minimum. The different optimized structures and the MP2/AVTZ and CCSD(T)/AVTZ (in parenthesis) VDEs are shown in Figure \ref{OHcluster}. The latter have been calculated at the MP2/AVTZ optimized geometries.
\textsc{
\begin{center}
\begin{figure}[]
\centering
\includegraphics[scale=0.5]{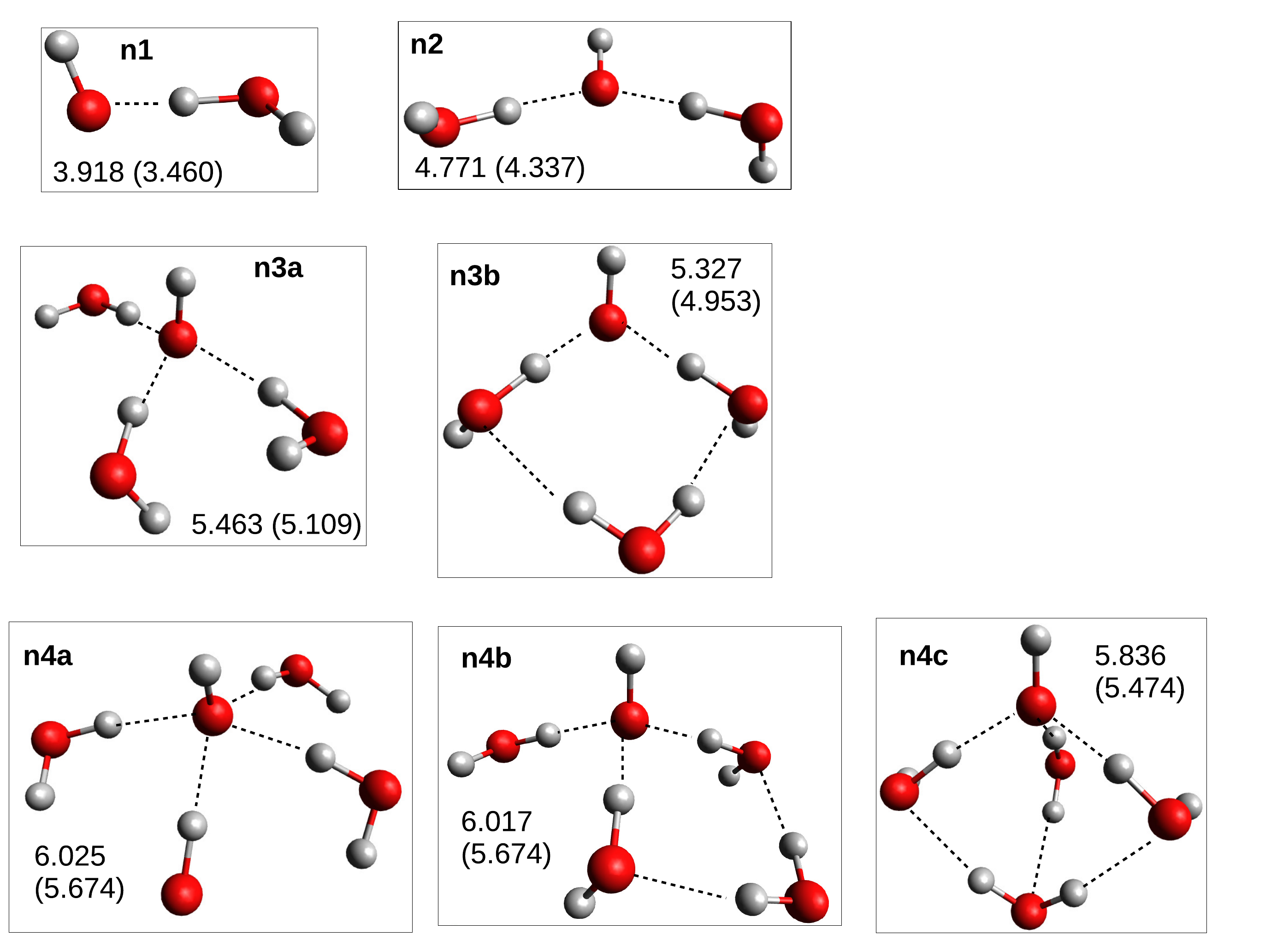}
\caption{MP2/AVTZ global minima for the hydrated hydroxide clusters OH(H$_{2}$O)$_{n}^{-}$ for n=1 to n=4. MP2/AVTZ and CCSD(T)/AVTZ (in parenthesis) vertical detachment energies in eV.}
\label{OHcluster}
\end{figure}
\end{center}
}   
\noindent The MP2 and CCSD(T) energies lead to the same most stable structures for $n=3$ and $n=4$ which corresponds to the $n3b$ and $n4c$ isomers, respectively. Lee et al. \cite{Lee2004} obtained different results using the same methodology (MP2 optimization followed by CCSD(T) single point energy) with a slightly more diffuse basis set. However, no change in the stability order of the cluster isomers were seen when using AVQZ basis set or extending diffuse functions. We did not compute the activation barrier related to the different isomerization pathways from one isomer to the other. However, since it involves several bonds breaking, we expect them to be very high. This allows to treat the different isomers as separate species. Additional calculations and discussions can be found in the supplementary material.  \\ 
As can be seen, the VDE increases with the size of the cluster. This is not surprising since the negative charge is stabilized by the water shell \cite{Morita2014}. This has also been shown for the water cluster anions (H$_{2}$O$)_{n}^{-}$ \cite{Kim1999}. However, an interesting difference between water cluster anions and hydrated hydroxide anions can be pointed out: while the former are dipole-bound states with the excess electron situated on a very diffuse orbital, the latter are closed-shell species with more important correlation effects. Indeed the MP2 and CCSD(T) results are very different, with the former overestimating the VDE by 0.37 eV on average. Two reasons can be evoked: i) contribution from quadruple excitation terms from the closed $\pi$-like highest occupied molecular orbital (HOMO) that has been shown to be important in OH$^{-}$ \cite{Martin2001} and ii) the well known overestimation of the stability when using MP2 \cite{Tsuzuki2000,Wintjens2003}. Note that the difference in the interaction that bounds the excess electron in both water and hydroxide anion cluster leads to larger VDE for the latter. \\

\FloatBarrier

\section{Reactivity with H}
\label{sec_H}

In the present section, the H-OH(H$_{2}$O)$_{n}^{-}$ collision complex is investigated. \\

\subsection{H+OH(H$_{2}$O)$^{-}$}
\label{sec_Hn1}
  
\begin{figure}[]
\centering
\begin{subfigure}{0.5\textwidth}
\centering
\includegraphics[width=1\textwidth]{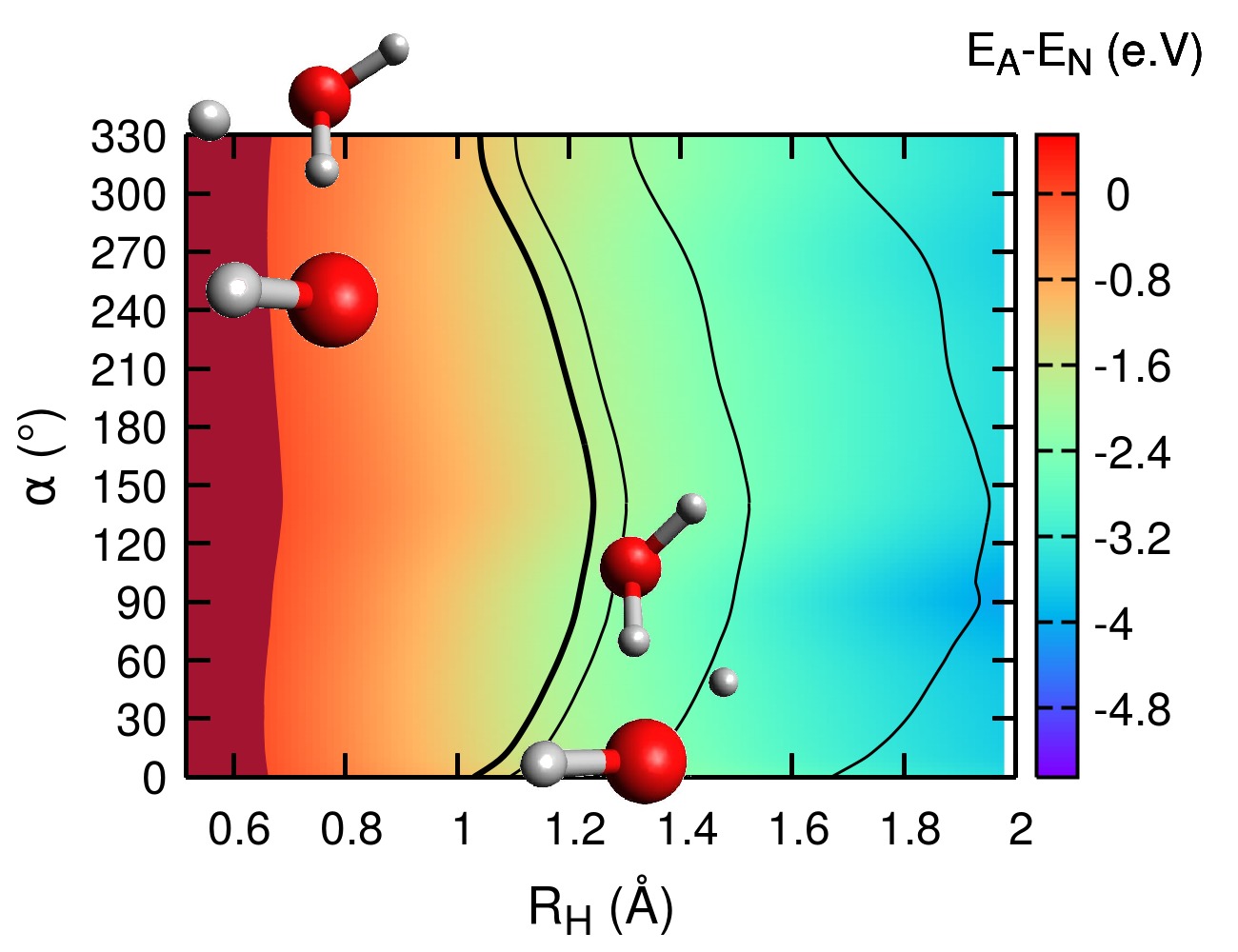}%
\caption{$\beta=0\degree$}
\label{H_bet0}
\end{subfigure}%
\begin{subfigure}{0.5\textwidth}
\centering
\includegraphics[width=1\textwidth]{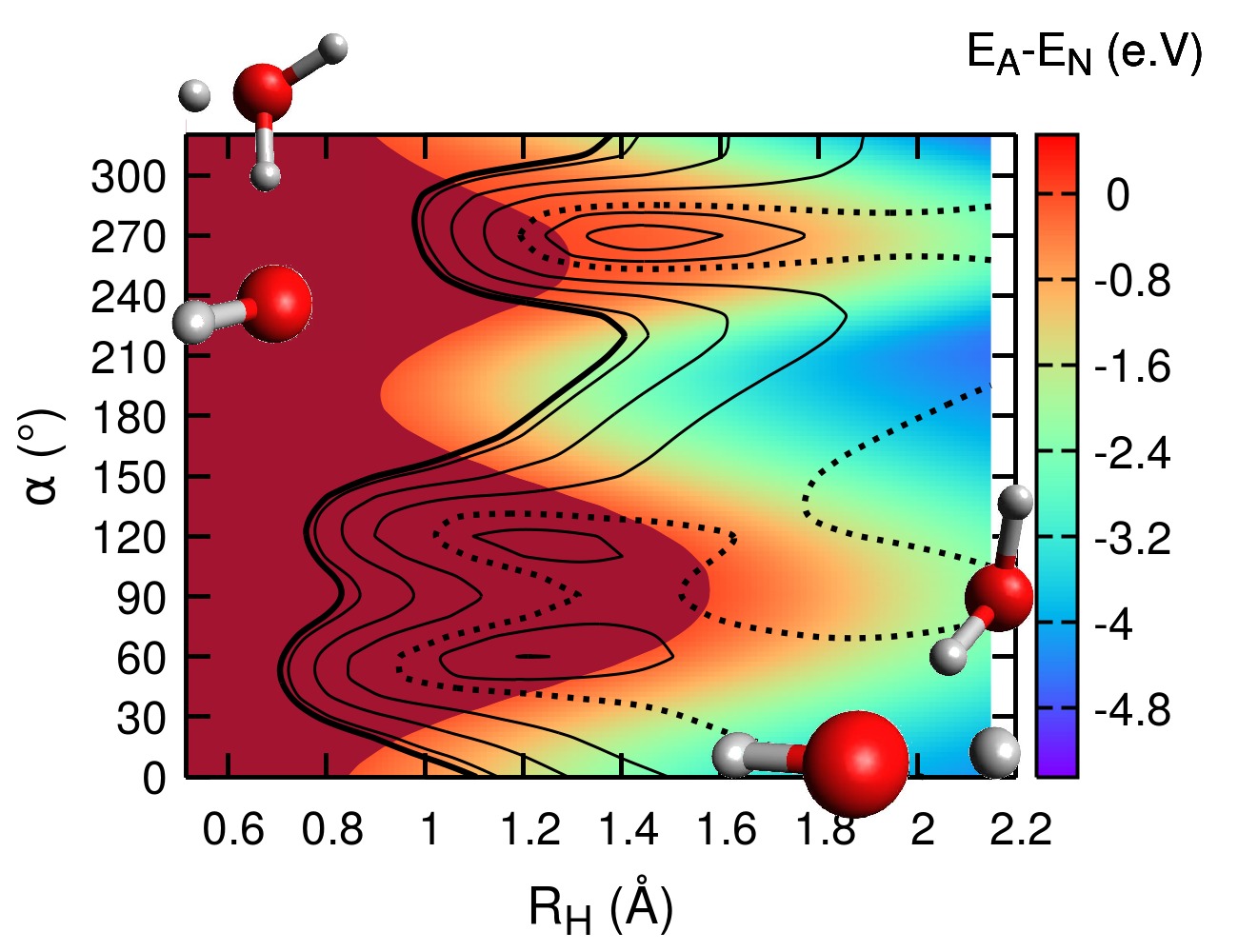}%
\caption{$\beta=45\degree$}
\label{H_bet45}
\end{subfigure} \\
\begin{subfigure}{0.5\textwidth}
\centering
\includegraphics[width=1\textwidth]{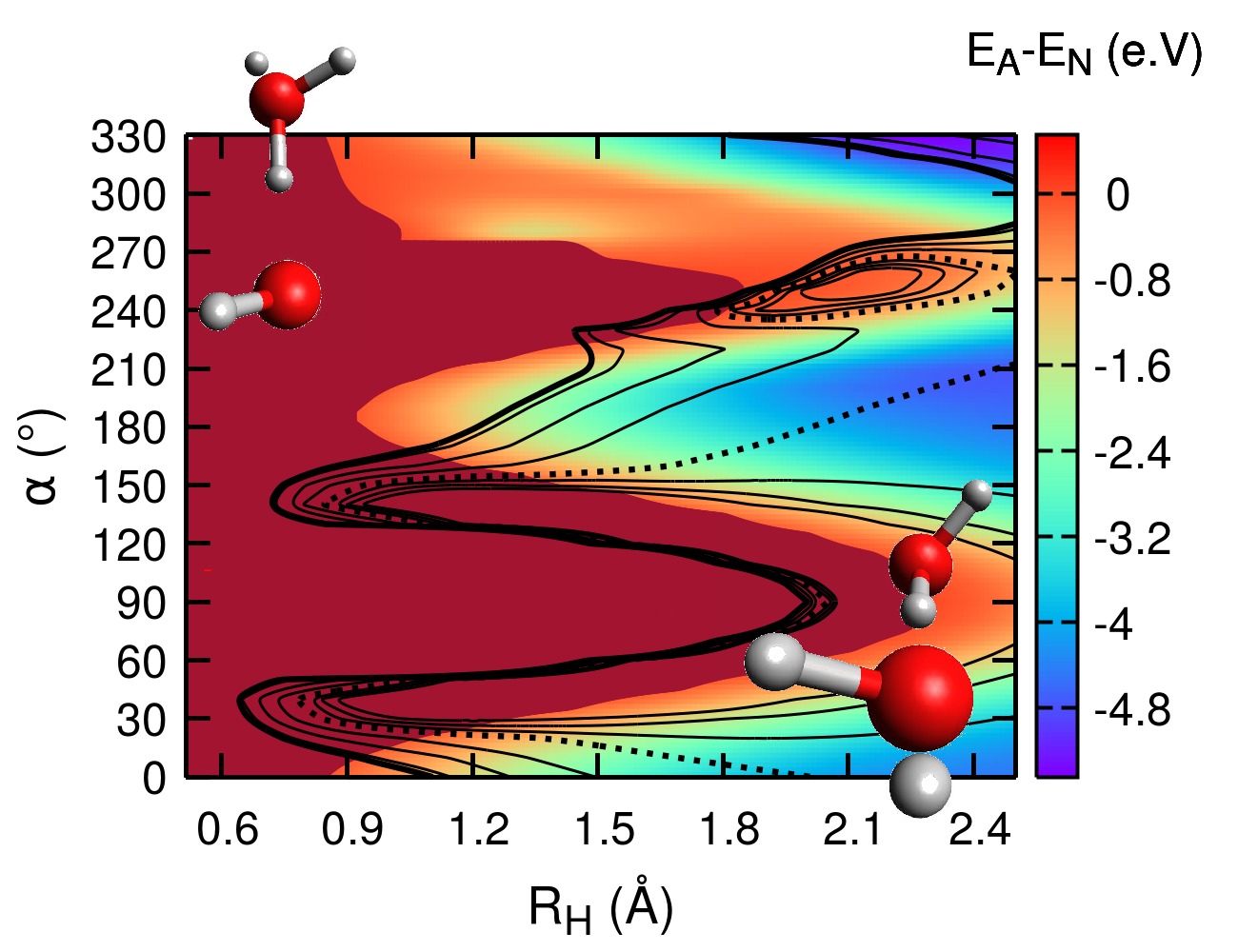}%
\caption{$\beta=90\degree$}
\label{H_bet90}
\end{subfigure}%
\begin{subfigure}{0.5\textwidth}
\centering
\includegraphics[width=1\textwidth]{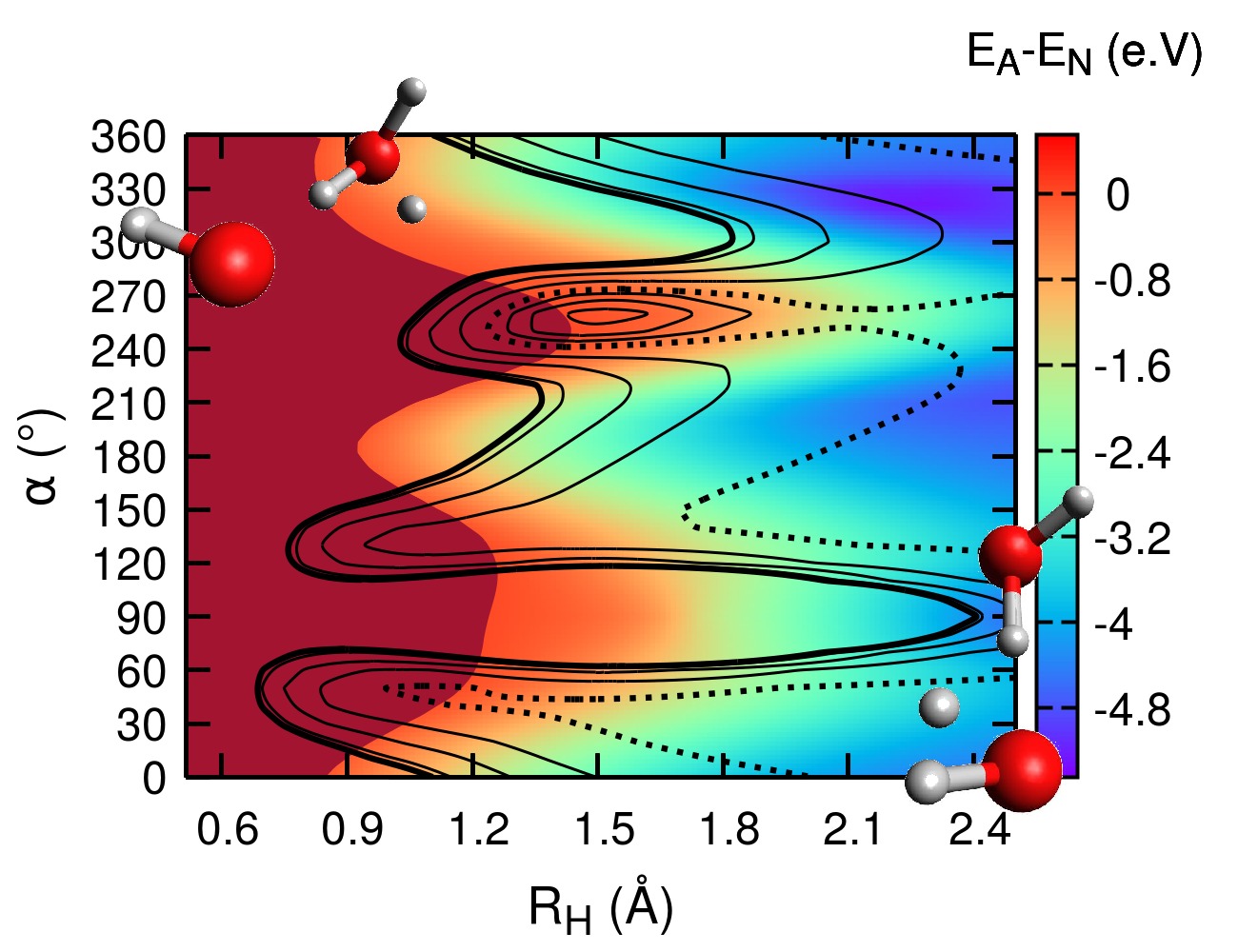}%
\caption{$\beta=135\degree$}
\label{H_bet135}
\end{subfigure}%
\caption{MP2/AVTZ/AVQZ cuts of the energy difference between anion H-OH(H$_{2}$O)$^{-}$ and neutral H-OH(H$_{2}$O) potential energy hyper-surfaces. The Jacobi coordinate $\alpha$ has been plotted against $R_{H}$ for specific values of $\beta$. The dark red region corresponds to the AD region where both PES crosses. The isolines represent the interaction energy of the anion where the thick dashed lines corresponds to the energy at dissociation (hence H+OH(H$_{2}$O)$^{-}$) and the thick line is E$_{0}$ the energy at the entrance channel (see text for details). The thin isolines are separated by 0.27 eV and 0.14 eV for the positive and negative values, respectively. The molecular cartoons correspond to geometries where $\alpha=90 \degree$ (bottom right for each figure) and $\alpha=260 \degree$ (top left for each figure).}
\label{Map_Hn1}
\end{figure}

Since we are dealing with a collision process, the Jacobi coordinate have been used to define the position of the collisional H atom in the interaction complex H-OH(H$_{2}$O)$^{-}$. The system of coordinates is illustrated in Figure \ref{n1_coord}.
\textsc{
\begin{center}
\begin{figure}[]
\centering
\includegraphics[scale=0.5]{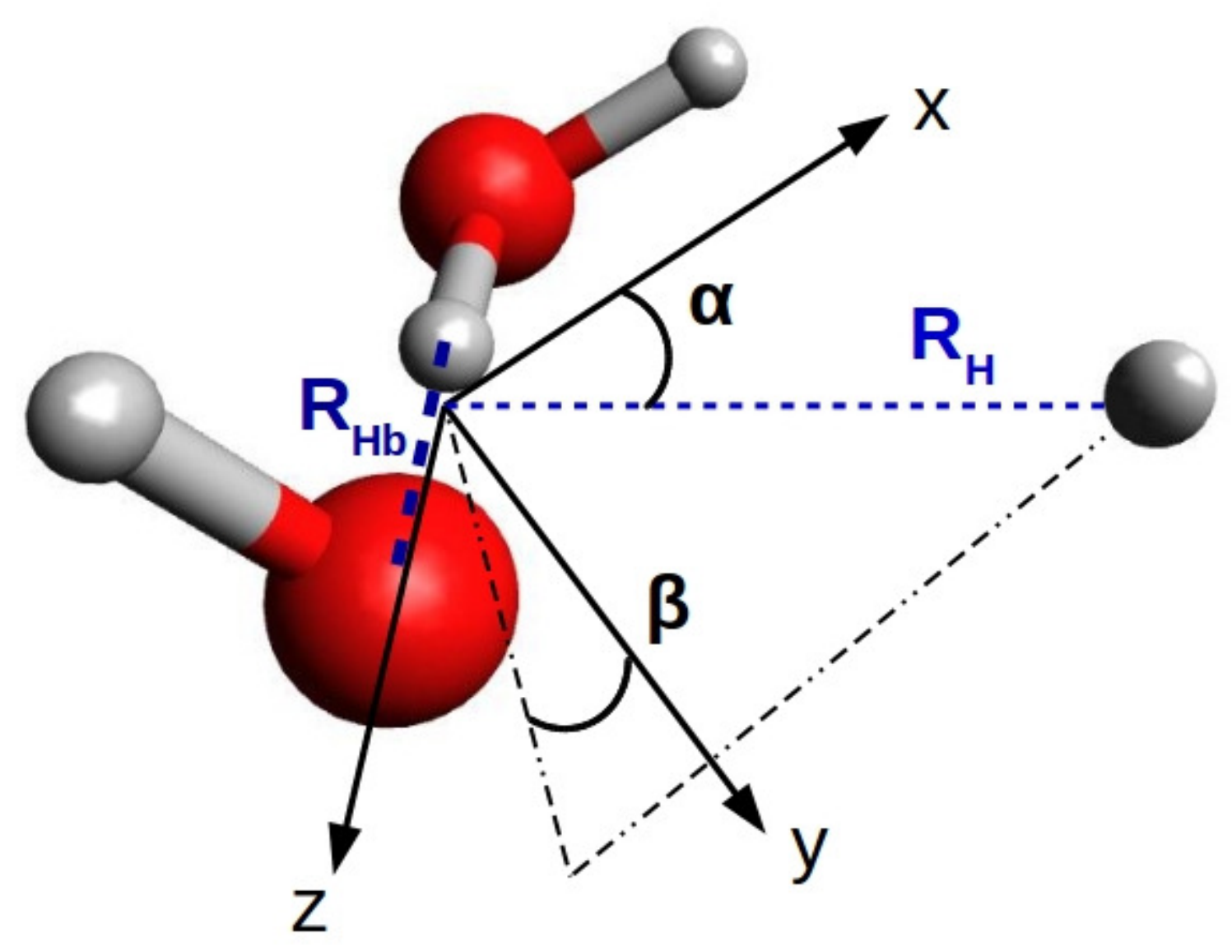}
\caption{Jacobi coordinate of the H-OH(H$_{2}$O)$^{-}$ molecular system.}
\label{n1_coord}
\end{figure}
\end{center}
}   
\noindent $R_{Hb}$ is the water bridge length, \textit{i.e} the distance between the O atom of the hydroxide group and the nearest H atom of the water group, $R_{H}$ is the distance between the centre of mass of OH(H$_{2}$O)$^{-}$ and H, $\alpha$ is the polar angle between the $x$ axis and the $R_{H}$ vector and $\beta$ is the azimuthal angle between the $y$ axis and the projection of the $R_{H}$ vector onto the $zy$ plane. Note that the $z$ axis has been chosen to coincide with $R_{Hb}$ while the $x$ axis is taken to be parallel to the second O-H bond of the water group. The internal coordinates of the OH(H$_{2}$O)$^{-}$ group have been kept frozen. 
The autodetachment region has been investigated along four collision angles: $\beta=0 \degree, 45 \degree, 90 \degree$ and $135 \degree$. In order to represent it in a comprehensive way we have plotted in Figure \ref{Map_Hn1} the energy difference between the anion and neutral PES as a function of R$_{H}$ and $\alpha$. The regions where this value becomes positive, \textit{i.e} when the neutral PES is below the anion's, are highlighted in red.
Furthermore, to see if the AD region can be reached, we have plotted several isolines of the anionic PES: the energy at dissociation (0 eV) is shown in dashed line while the effective total energy in the entrance channel at $T$=300 K is shown in thick solid line. The latter is given by E$_{0}=$E$_{\textrm{tr}}$+E$_{\textrm{rot}}$+E$_{\textrm{vib}}$ where each term is related to the translational, rotational and vibrational energy at the entrance channel. To get an estimation of E$_{0}$, we have used the classical average thermal energy for E$_{\textrm{tr}}$ and E$_{\textrm{rot}}$; $k_{B}$T and $3/2k_{B}T$, respectively. We thus assume an equipartition of the energy among the translational and rotational degrees of freedom. E$_{\textrm{vib}}$ is taken as the sum of ZPEs associated to the normal mode of OH(H$_{2}$O)$^{-}$, obtained from MP2/AVTZ frequency calculations, yielding E$_{\textrm{vib}}=0.849$ eV. Note that this means that we assume all vibrational modes to be in their respective ground state, our E$_{\textrm{vib}}$ value thus corresponds to a lower limit. The final value for E$_{0}$ at 300 K is 0.912 eV. As can be seen in Figure \ref{Map_Hn1}, the AD region is outside the energy range for $\beta=0 \degree$ (top left panel). However, the AD region may be reached for different collision angles. In particular, the AD region lies lower in energy for geometries where the H atom approaches the O atom of the OH group. \\
By examining the isolines, which correspond to the anion PES, one can also make some interesting observations. The interaction potential between H and OH(H$_{2}$O)$^{-}$ is attractive at large R$_{H}$ distance due to the charge-induced dipole electrostatic interaction but it is highly anisotropic at intermediate distances. Indeed, the interaction energy is almost repulsive when approaching the H atoms from both the hydroxide and water group. The top left panel in Figure \ref{Map_Hn1} corresponds to a collision on a plane perpendicular to the H-bond and is therefore almost flat with respect to the $\alpha$ angle. \\  
To better support our observations for both the interaction potential of the anion and the AD region, we have also computed potential energy curves for three different configurations: i) with the H atom approaching the cluster from the H side of the OH group, ii) with H approaching the O atom of the water group and iii) with H approaching from the O atom of the hydroxide group. The MP2 and CCSD(T) method with AVTZ/AVQZ basis set are compared. Furthermore, since water cluster anions are dipole bound species, we have also studied the effect of additional diffuse functions. The details of these calculations are provided in supplementary material. They confirm that the configuration where the H atom approaches the O from the OH group corresponds to the "preferred" collisional path. In addition, we pointed out that the addition of diffuse functions stabilises the anion by about 0.2 eV in this configuration, leading to a larger VDE and hence a higher lying AD region, while having only little impact for other collisional approaches. \\     
So far, all calculations have been performed in the frozen cluster approximation. Since relaxation effects may influence the VDE, we have investigated the collision process from the preferred approach, corresponding to the H atom colliding from the O side of the OH group. A MP2/AVTZ optimization on the Jacobi coordinates with the internal coordinates of the cluster kept frozen yields the following values: R$_{H}=1.88 \angstrom, \alpha=65 \degree$ and $\beta=78 \degree$. A subsequent relaxed geometry optimization was performed by unfreezing the cluster \textit{i.e} all coordinates are optimized. Both structures can be seen in Figure \ref{Hn1_geo} along with their MP2 and CCSD(T) VDE.
\textsc{
\begin{center}
\begin{figure}[]
\centering
\includegraphics[scale=0.4]{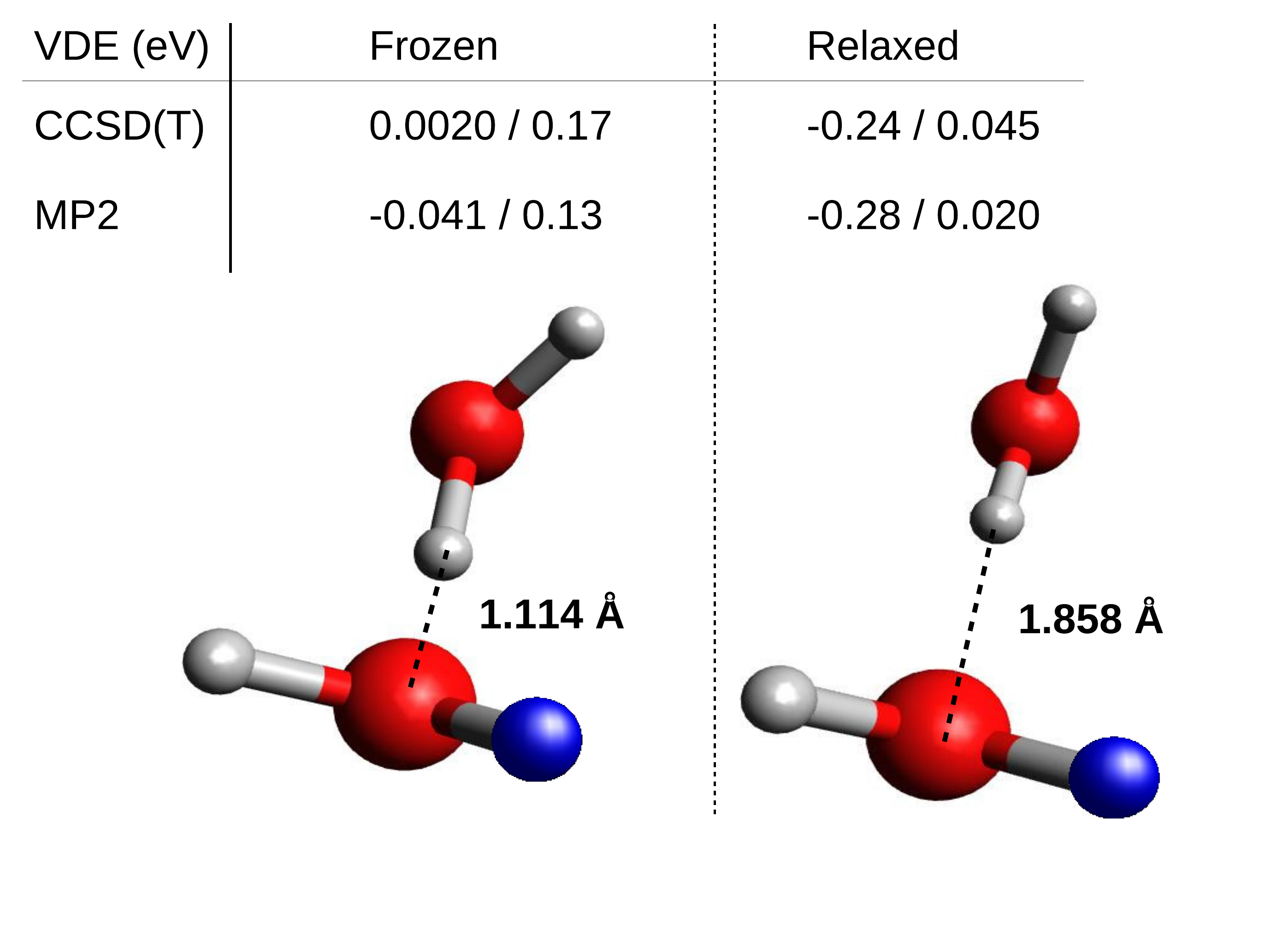}
\caption{MP2/AVTZ geometry optimization of the H-OH(H$_{2}$O)$^{-}$ molecular system. Left figure: frozen cluster, right figure: relaxed cluster. MP2 and CCSD(T) vertical detachment energy (VDE) with the AVTZ/AVQZ (left values) and AVTZ/AVQZ+aug basis set (right values). The collisional H atom is shown in blue.}
\label{Hn1_geo}
\end{figure}
\end{center}
}   
\noindent The relaxation of the inner coordinate of the OH(H$_{2}$O)$^{-}$ group leads to a lengthening of the R$_{Hb}$ distance and a rotation of the H$_{2}$O group where the optimized geometry corresponds to the water dimer anion \cite{Chen1999}. The energy is stabilized by about 1.2 eV while the VDE decreases remarkably. In addition, the effect of diffuse functions on the latter is rather strong which, again, demonstrates the dipole-bound nature of the HOMO. If the collision is fast, the cluster will not have time to relax (diabatic case) but at low collision energies, the relaxation may have sufficient time to occur (adiabatic case). This will have a strong impact on the position of the AD region. However, the VDE remains very small, even for the unrelaxed (frozen) case. Based on the above discussions we can conclude that direct detachment mechanism leading to the HOH(H$_{2}$O)+$e^{-}$ exit channel will only be possible for higher collision energies and/or larger internal temperature of OH(H$_{2}$O)$^{-}$. Moreover, the ejected electron would need to carry enough energy to prevent the water group from dissociating. Based on the position of the AD region, a direct mechanism for the AED reaction will only be possible at high collision energy and/or accounting for higher excited ro-vibrational states. Therefore, a indirect mechanism, leading to detachment through water loss, is more likely to explain the AED reaction. Additional calculations, confirming the later assumption are provided in supplementary material. Assuming that the electron quickly autodetach when the water anion H$_{2}$O$^{-}$ is formed, the rate of the AED reaction will depend on the rate of water loss.

\FloatBarrier

\subsection{H+larger clusters}
\label{sec_Hn}

In addition to the steric effect of water groups, the increase of the VDE for increasing $n$ should also be taken into account to explain the smaller measured AED reaction's rate constant for larger clusters \cite{Howard1975}. In particular, the autodetachment region for the collisional complex H+OH(H$_{2}$O)$_{n}^{-}$ will lie higher in energy even for minimum collision angles (where H approaches the O atom of the OH group). This is due to the excess charge being de-localized for increasing size of the cluster. As already pointed out above, the preferred collision angle, which correspond to the H atom approaching the O atom of the OH group, also corresponds to the lower VDE, \textit{i.e} where the AD region is most easily reached. \\
We first performed geometry optimization on the Jacobi coordinates with the initial geometry corresponding to H facing O of the OH group. The internal coordinates of the cluster have been kept frozen. We then scanned the interaction energy over the R$_{HO}$ distance (distance between H and O). The relevant geometries can be seen in Figure \ref{EAEint_Hcluster} along with their corresponding VDE and interaction energies. The effect of additional diffuse functions as well as of the method has been investigated. 

\textsc{
\begin{figure}[!htb]
    \centering
    \begin{tabular}[t]{cc}
\begin{subfigure}{0.5\textwidth}
    \centering
    \includegraphics[width=1\textwidth]{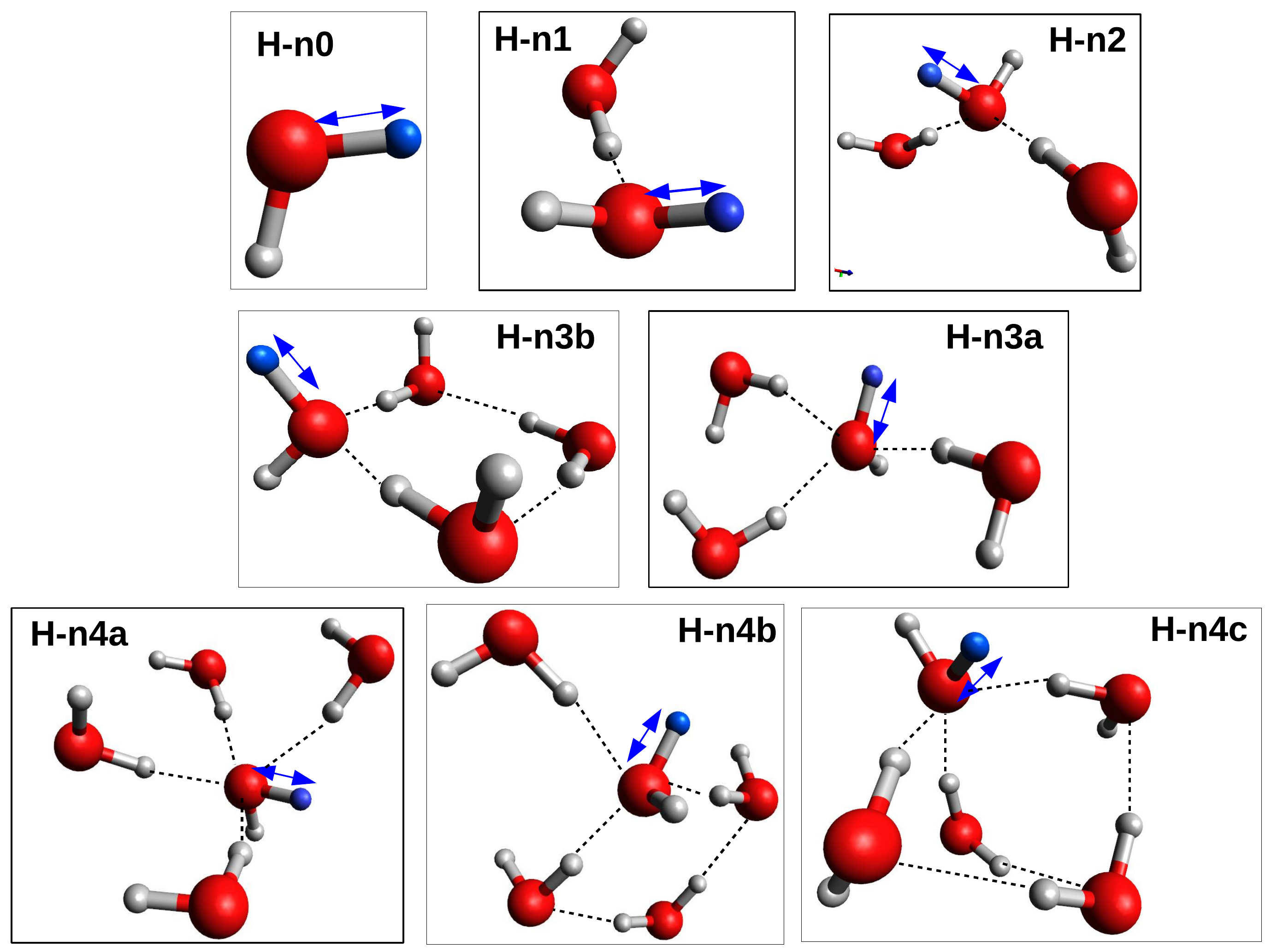}
\end{subfigure}
    &
        \begin{tabular}{c}
            \begin{subfigure}[t]{0.45\textwidth}
                \centering
                \includegraphics[width=1\textwidth]{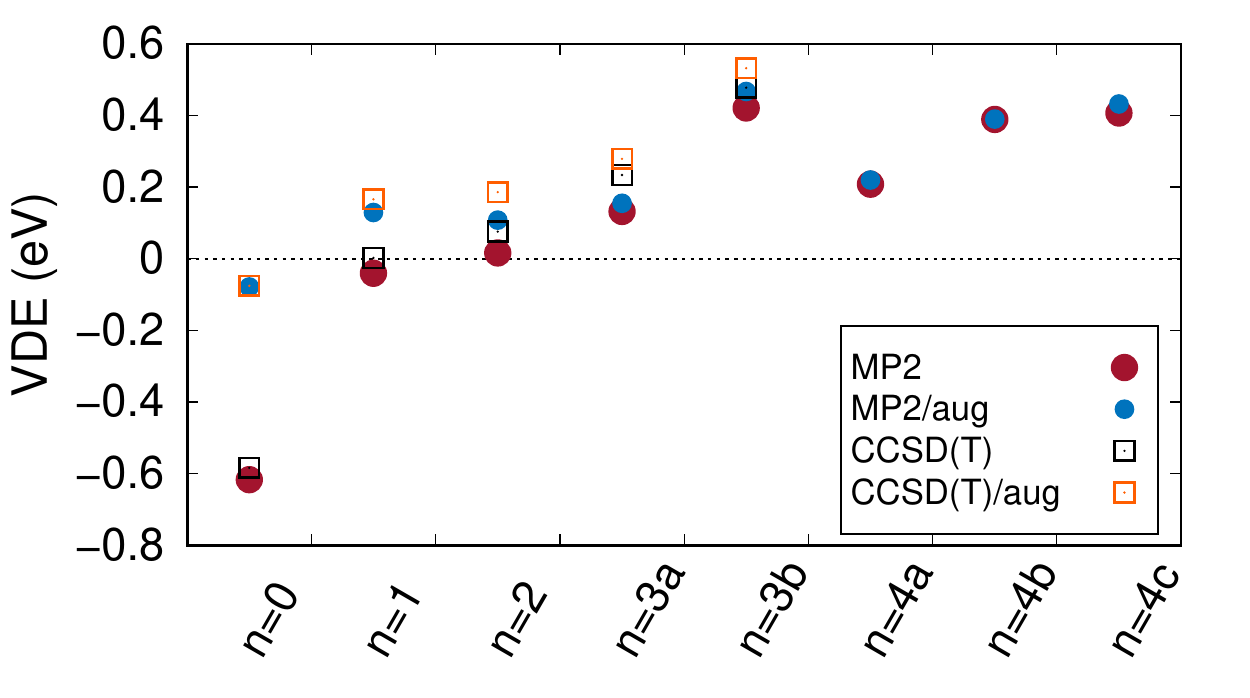}                
            \end{subfigure}\\
            \begin{subfigure}[t]{0.45\textwidth}
                \centering
                \includegraphics[width=1\textwidth]{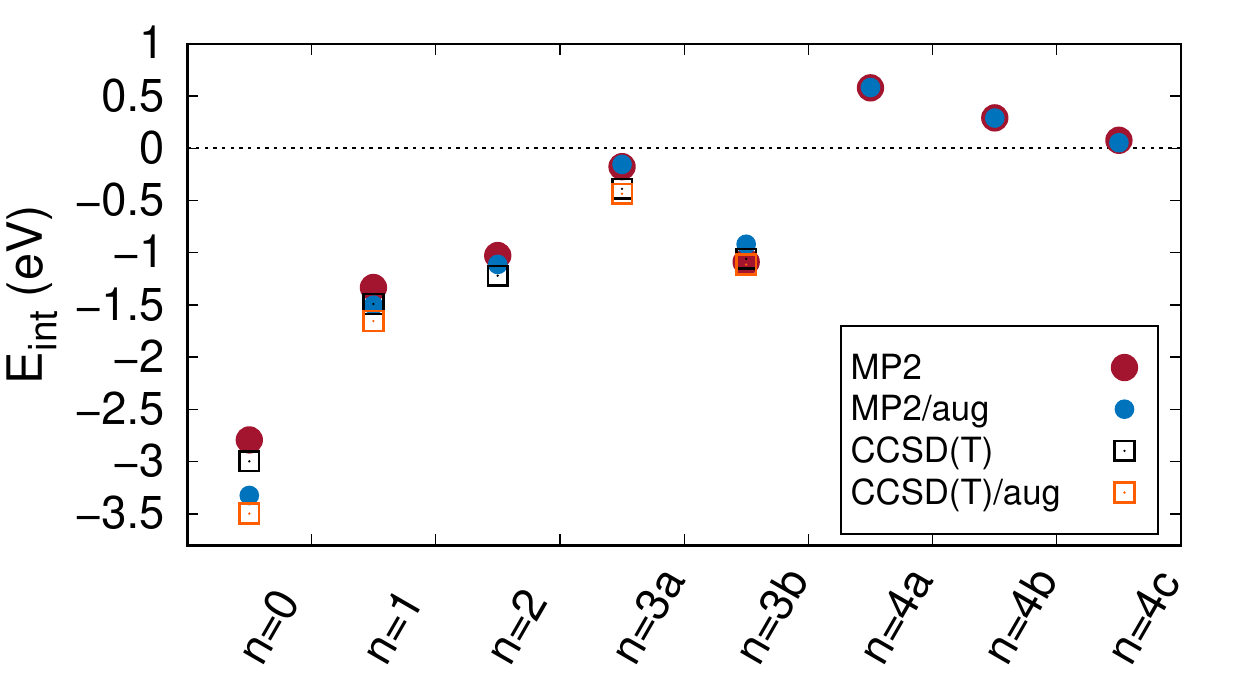}
            \end{subfigure}
        \end{tabular}\\
    \end{tabular}
    \caption{VDE and interaction energy at the preferred approach configuration for the H-OH(H$_{2}$O)$_{n}^{-}$ collisional complexes. The geometries are shown on the left picture and have been obtained by optimizing the Jacobi coordinates of the incoming H atom (shown in blue) at the MP2/AVTZ/AVTZ level of theory. The VDE and interaction energies have been calculated on the obtained geometries using MP2 and CCSD(T) method (except for n=4) with AVTZ/AVQZ and AVTZ/AVQZ+aug basis set. The BSSE has been taken into account for E$_{int}$.}  
\label{EAEint_Hcluster}
\end{figure}
}

\noindent The VDE increases with the size of the cluster which is, like for OH(H$_{2}$O)$_{n}^{-}$, due to the stabilization of the excess electron. For the same reason, the interaction energy decreases for larger clusters. In particular, E$_{\textrm{int}}$ is stronger for the isomer where the reaction site (O atom of the OH group) is more accessible ($n3b$ and $n4c$), suggesting isomer-dependent reaction rates. For $n=0$ an $n=1$ the addition of diffuse functions has a stronger effect on both the VDE and the interaction energy than the method. This underlines the dipole-bound nature of the clusters. For larger clusters, $n=3$ and $n=4$, the situation is opposite \textit{i.e} the method, hence the extent of electron correlation, is more important. The $n=2$ cluster seems to be an intermediate case. This observation leads to the conclusion that the dipole-bound nature of the HOMO decreases for increasing size of the cluster where correlation effects become more important. This is also seen for the interaction energy. The latter globally decreases with increasing $n$ and becomes positive for $n=4$. However, since a minimum has been found, the potential is not repulsive. Isomer-dependent effects can be seen. In particular, the $n3a$ and $n4a$ clusters exhibit the smallest $E_{int}$ and VDE. Two reasons can be put forth: i) the higher coordination number for $n3a$ and $n4a$, which ultimately decreases the partial charge on O and, hence, the O-H bond strength, and ii) the more exposed hydroxide group which is thus more easily attainable while being more effectively "screened" for $n4c$ and $n3b$. This result underlines the sensitivity of the incoming H atom to its chemical environment. \\       
In order to investigate the accessibility of the AD region, we have calculated the anion and neutral potential energy curve (PEC) of H+OH(H$_{2}$O)$_{n}^{-}$ (for $n=1,2,3,4$) corresponding to the preferred approach. The results for $n=1,2,3b$ and $4c$ are shown in Figure \ref{Cross_H}.
\textsc{
\begin{figure}[]
\centering
\begin{subfigure}{0.6\textwidth}
\includegraphics[width=1\textwidth]{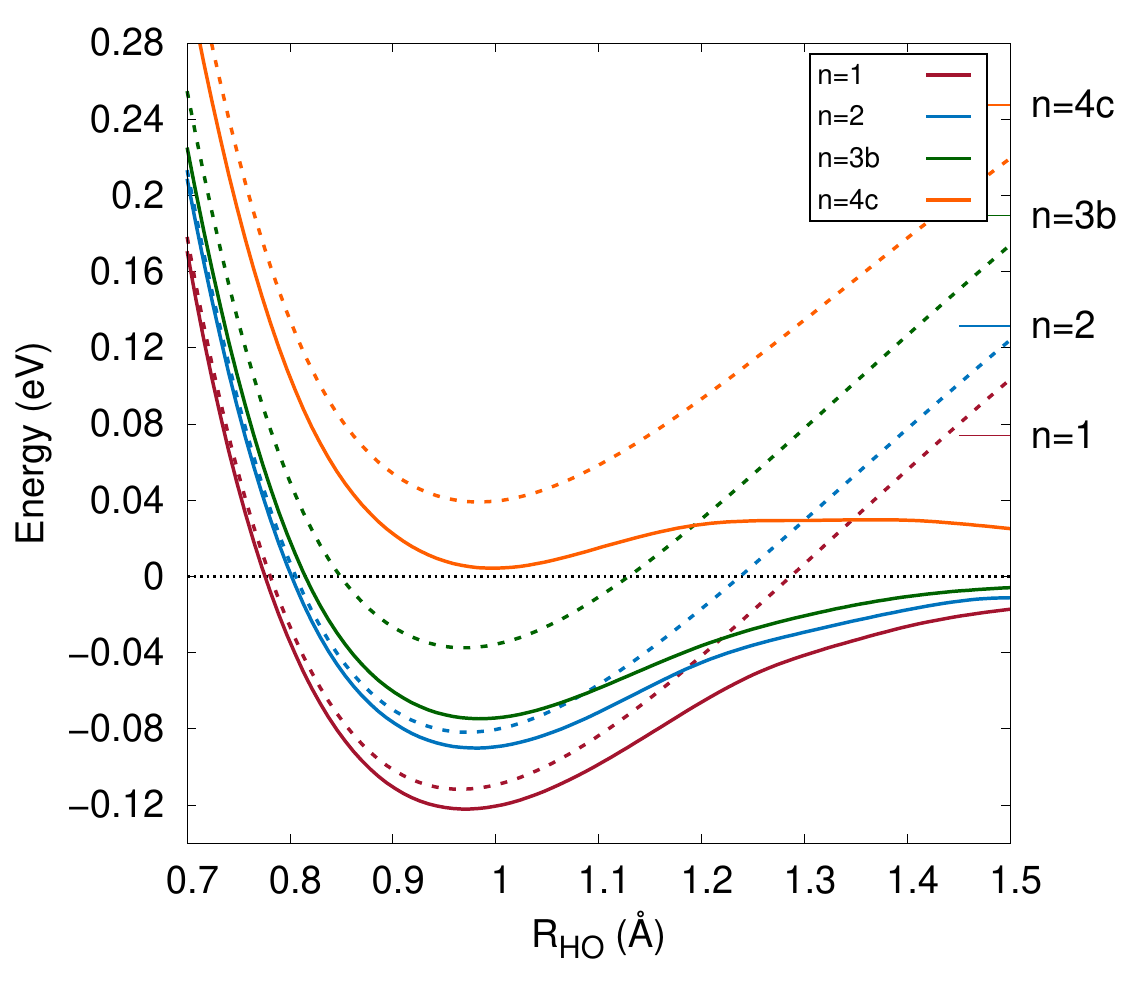}
\end{subfigure}
\caption{Preferred approach MP2/AVTZ/AVQZ+aug potential energy curves for the anion (solid lines) and neutral (dashed lines) H+OH(H$_{2}$O)$_{n}^{-}$ collisional system. The black dashed line corresponds to the 0 energy while the coloured horizontal lines correspond to the energy at the entrance channels E$_{0}$ (see text for details) for each cluster.}  
\label{Cross_H}
\end{figure}
}   
The autodetachment region lies higher energy for increasing size of the cluster. In particular, it becomes hardly reachable for $n=3$ and $n=4$. Even though our first results for $n=1$ (see Figure \ref{Map_Hn1}) show that the direct detachment is feasible for some collisional angles, the inclusion of diffuse functions shifts the AD region to higher energies. This implies that the direct detachment process would only be possible for small clusters ($n=1$ and $n=2$) for excited vibrational states or at higher collision energies. The indirect mechanism would be needed to explain the detachment process for collisions between H and larger clusters and/or at low temperature. The most stable isomers, \textit{i.e} $n3b$ and $n4c$, also exhibit the strongest interaction energies, largest VDE (see Figure \ref{EAEint_Hcluster}) and, consequently, the higher lying AD region. The comparison between isomers PEC are given in supplementary material. Our results show that the height of the barrier found along the $R_{HO}$ coordinate is 0.8 eV, 0.52 eV and 0.37 eV for $n4a$, $n4b$ and $n4c$, respectively. Furthermore, a potential barrier, not present for $n3b$, appears for the $n3a$ isomer with a height of around 0.27 eV. This observation supports the hypothesis of isomer-dependent rate constants.
In order to consider the indirect mechanism , we have chosen to investigate the $n3a$ cluster as a test case. Figure \ref{Hn3_RHb} shows the potential energy surface of the H-OH(H$_{2}$O)$_{n}^{-}$ species, obtained at the MP2/AVTZ/AVTZ level of theory by scanning the R$_{Hb1}$ and R$_{Hb2}$ coordinates, which corresponds to the distance between the OH group and the two water groups, respectively. The position of the incoming H atom (coloured in blue) has been kept frozen at its preferred approach. The stability of the remaining fragment obtained for R$_{Hb1} \rightarrow \infty$ has been calculated using the MP2 and CCSD(T) method with AVTZ/AVTZ and AVTZ/AVQZ basis sets. The scanned map, relevant structures and coordinates as well as the calculated VDE are also depicted. All dissociation limits lie below the entrance channel energy since all dissociation and detachment channels are open, \textit{i.e} the reactions are all exothermic. The AD region has been highlighted in red and several isoenergetic lines of the anion PES are also shown. We have also carefully checked the charge distribution using Mulliken population analysis: the negative charge remains on the HOH(H$_{2}$O)$_{2}^{-}$ fragment. The anion and neutral PES crosses around R$_{Hb1}=10 \angstrom$, making the fragment unstable upon detachment with the VDE MP2/AVTZ/AVTZ value tending towards -0.049 eV at dissociation (large values of R$_{Hb1}$). As already discussed above, the calculated VDE is very sensitive to the method, basis set and relaxation effects. For the obtained H-OH(H$_{2}$O)$_{2}$ fragment, the calculated VDE values range from -0.023 to 0.040 eV. 
\textsc{
\begin{figure}[t]
\centering
\begin{subfigure}{0.4\textwidth}
\centering
\includegraphics[scale=0.6]{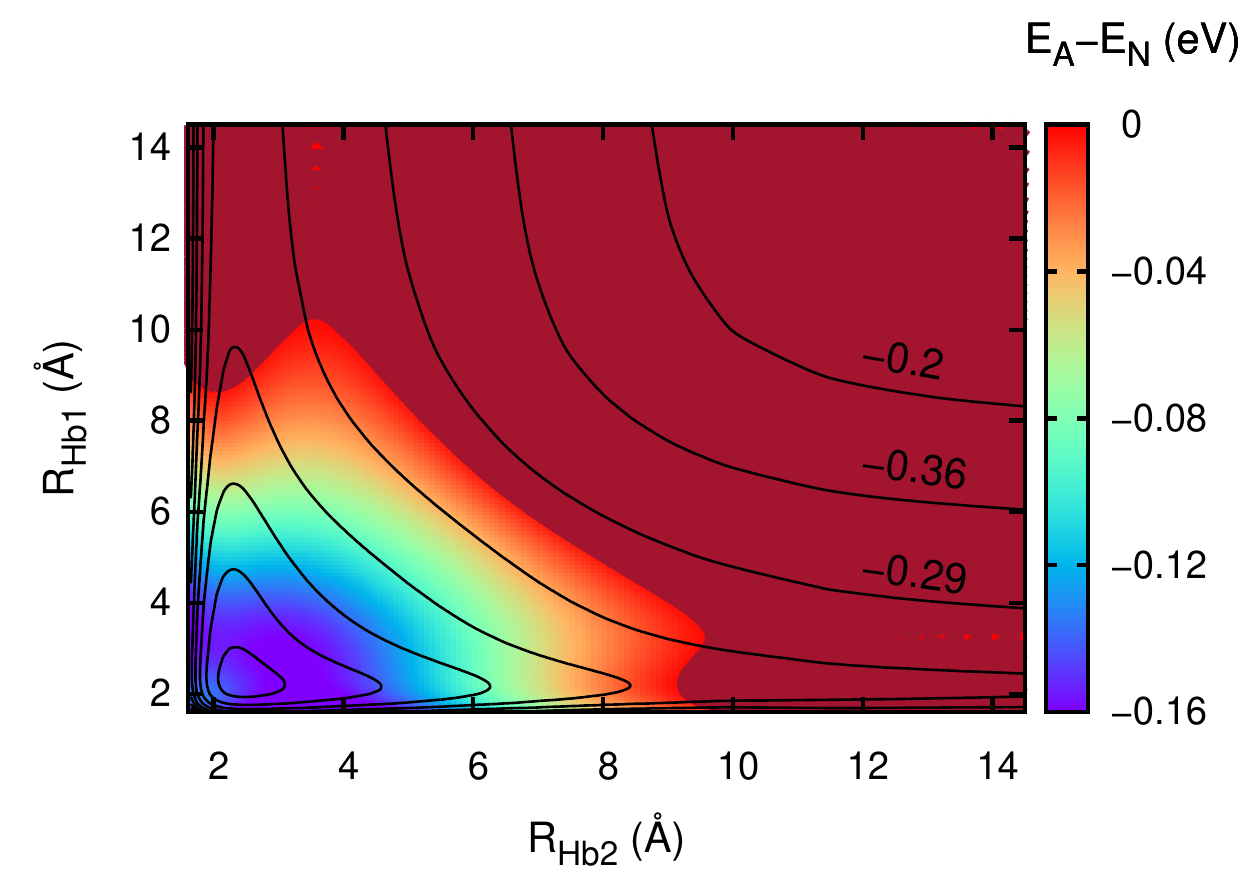}
\end{subfigure}
\hspace{35pt}
\centering
\begin{subfigure}{0.4\textwidth}
\centering
\includegraphics[scale=0.3]{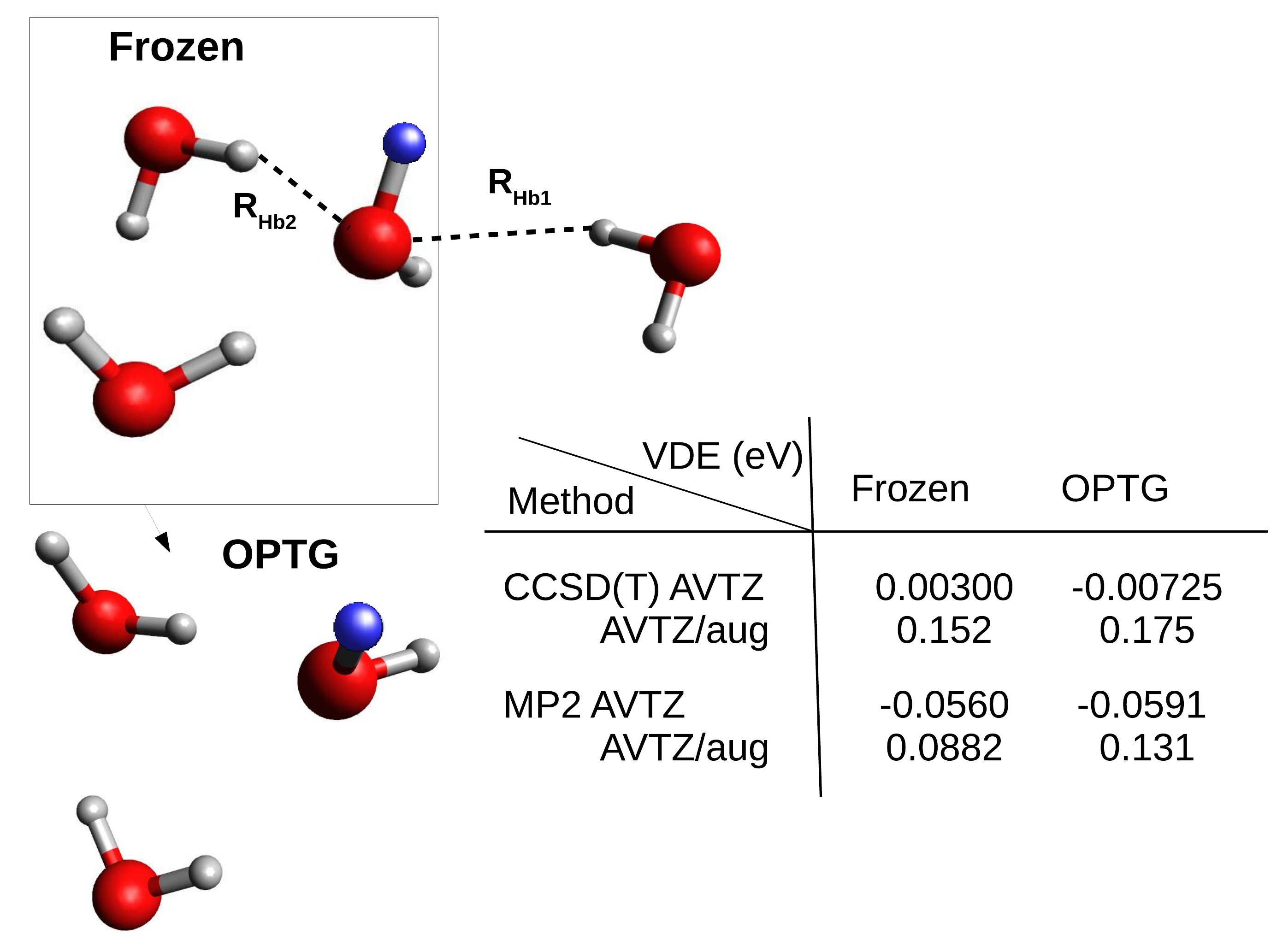}
\end{subfigure}
\caption{Right panel: difference between the anion and neutral MP2/AVTZ/AVTZ energy of the H-OH(H$_{2}$O)$_{3}^{-}$ specie along the R$_{Hb1}$ and R$_{Hb2}$ coordinates. The isoline corresponds to iso-contour of the anion PES with a step of 0.07 eV starting at -0.8 eV. The higher isoline is labelled. The energies are given relative to the entrance channel H+H(H$_{2}$O)$_{3}^{-}$. \\
Left: geometry of the investigated specie with the labelling of the R$_{Hb2}$ and R$_{Hb1}$ coordinates (top). MP2/AVTZ optimized geometry of the fragment obtained at large R$_{Hb1}$ distance. The calculated CCSD(T) and MP2 VDE are given in the table. The results in parenthesis have been obtained with AVQZ basis set.}
\label{Hn3_RHb}
\end{figure}
}   
Nevertheless the values are rather small, suggesting the production of unstable negatively charged fragments. Furthermore, the fragments will most likely be vibrationally excited, opening vibrational induced detachment channels which are expected to be important for water cluster anions \cite{Politzer1991}. This also supports the hypothesis that the detachment process occurs along the dissociation path. This confirms that indirect detachment mechanisms are likely to be dominant for collisions involving larger clusters. In addition, since the total fragmentation is exothermic for cluster size up to n=4, the produced fragment will eventually lead to water. If the excess electron has not been detached along the dissociation path, it will ultimately be ejected since the H$_{2}$O$^{-}$ water anion is unstable (VDE$<0$). The production of water cluster anions H-OH(H$_{2}$O)$_{l}^{-}$ will only be possible for n$>4$ since the total fragmentation becomes endothermic. However, it will also depends on their stability against vibrational induced detachment. 
      
\FloatBarrier

\section{Reactivity with Rb}
\label{sec_Rb}

In the present section, we investigate the interaction between Rb and the hydrated hydroxide anion clusters. Large differences in the structure and reactivity are expected between Rb and H. First of all the RbOH$^{-}$ anion is stable with a VDE around 0.2 eV whereas the water anion is unstable. Furthermore, considering the collision process Rb/H + OH$^{-}$, the AD region is only reached for a small angular space in the case of Rb \cite{Kas2017} while almost all collision angles lead to detachment for H \cite{Houfek2016}. In addition, the Rb-OH bond (about 3.5 eV \cite{Kas2017}) is weaker than the H-OH bond of water (around 5.1 eV \cite{Maksyutenko2006}) leading to endothermic dissociation reaction channels.  

\subsection{Rb+OH(H$_{2}$O)$^{-}$}
\label{sec_Rb_n1}

\begin{figure}[]
\centering
\begin{subfigure}{0.5\textwidth}
\centering
\includegraphics[width=1\textwidth]{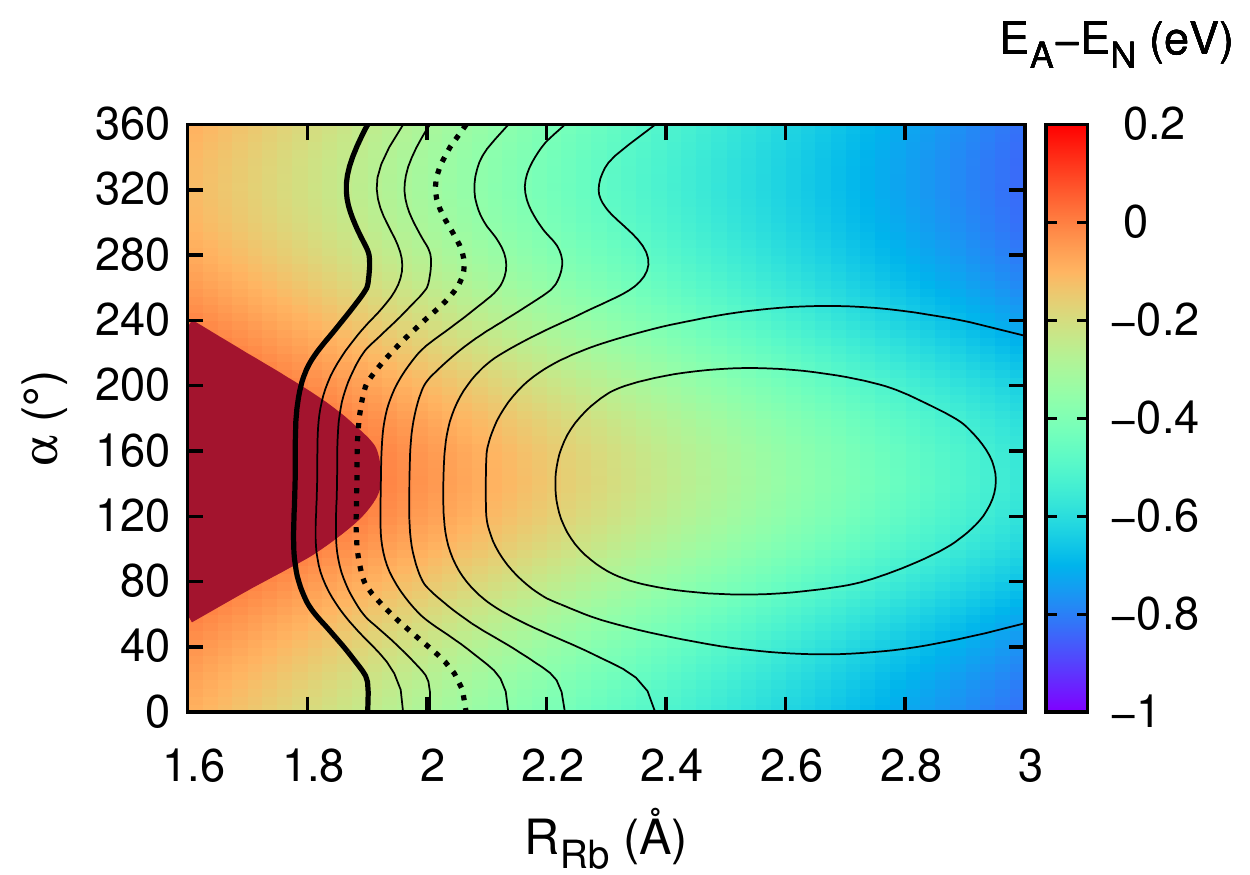}%
\caption{$\beta=0\degree$}
\label{Rb_bet0}
\end{subfigure}%
\begin{subfigure}{0.5\textwidth}
\centering
\includegraphics[width=1\textwidth]{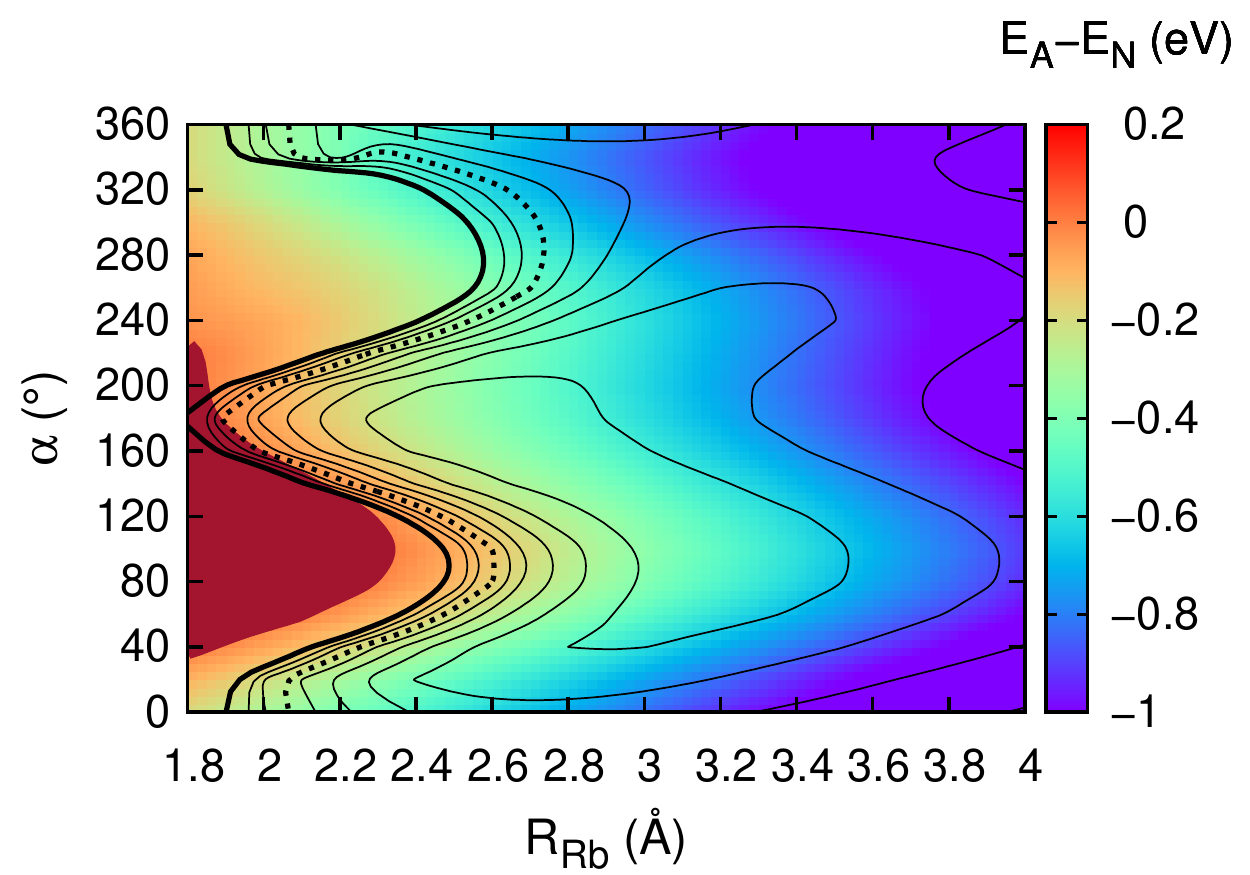}%
\caption{$\beta=45\degree$}
\label{Rb_bet45}
\end{subfigure} \\
\begin{subfigure}{0.5\textwidth}
\centering
\includegraphics[width=1\textwidth]{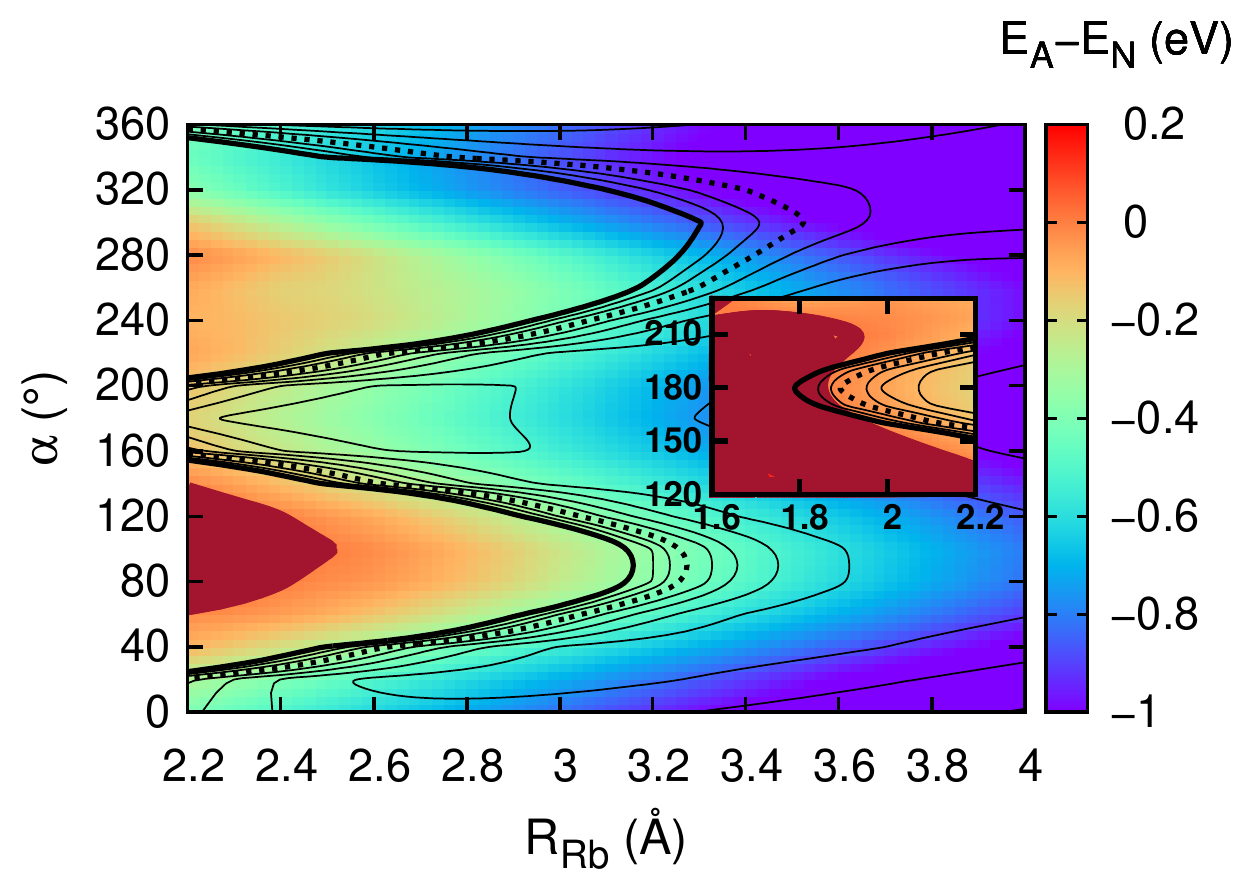}%
\caption{$\beta=90\degree$}
\label{Rb_bet90}
\end{subfigure}%
\begin{subfigure}{0.5\textwidth}
\centering
\includegraphics[width=1\textwidth]{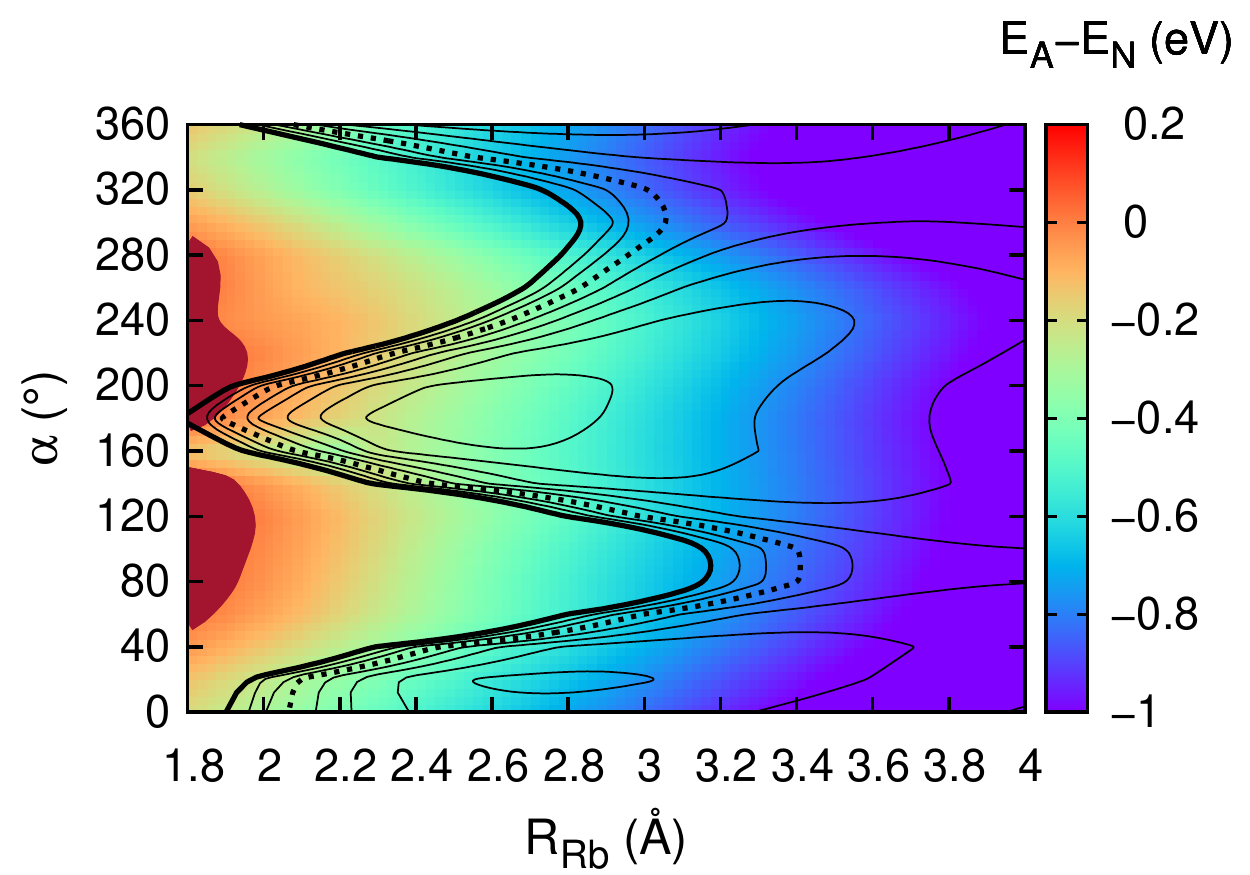}%
\caption{$\beta=135\degree$}
\label{Rb_bet135}
\end{subfigure}%
\caption{MP2/AVTZ/MDF$spdfg$ cuts of the energy difference between anion Rb-OH(H$_{2}$O)$^{-}$ and neutral Rb-OH(H$_{2}$O) potential energy hyper-surfaces. The Jacobi coordinates $\alpha$ has been plotted against $R_{Rb}$ for specific values of $\beta$. The dark red region corresponds to the AD region where both PES crosses. The isolines represent the interaction energies of the anion PES where the thick dashed lines corresponds to the energy at dissociation (hence Rb+OH(H$_{2}$O)$^{-}$) and the thick line is E$_{0}$ the energy at the entrance channel (see text for details). The thin isolines are separated by 0.27 eV and 0.14 eV for the positive and negative values, respectively.}
\label{Map_Rbn1}
\end{figure}

\begin{figure}[]
\centering
\begin{subfigure}{0.45\textwidth}
\includegraphics[width=1\textwidth]{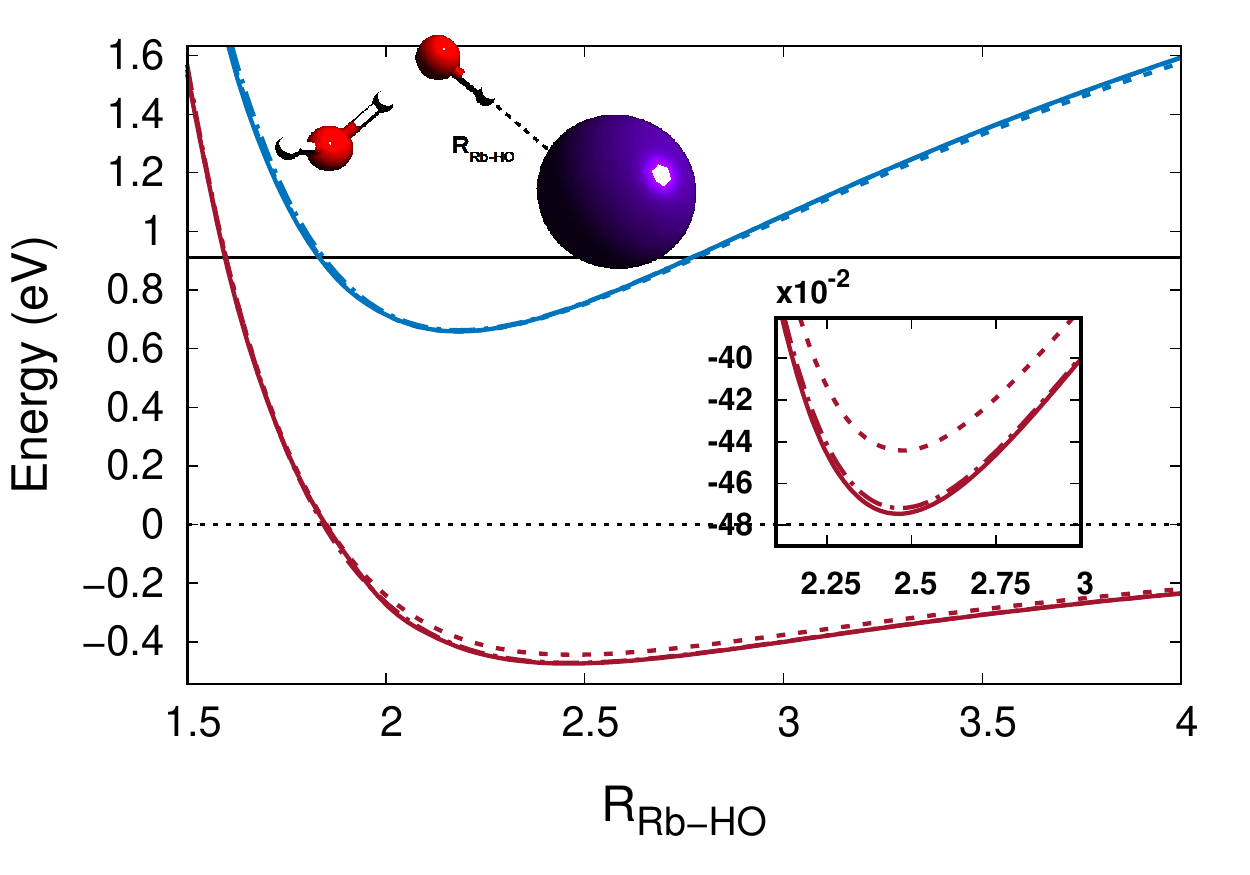}%
\end{subfigure}
\begin{subfigure}{0.45\textwidth}
\includegraphics[width=1\textwidth]{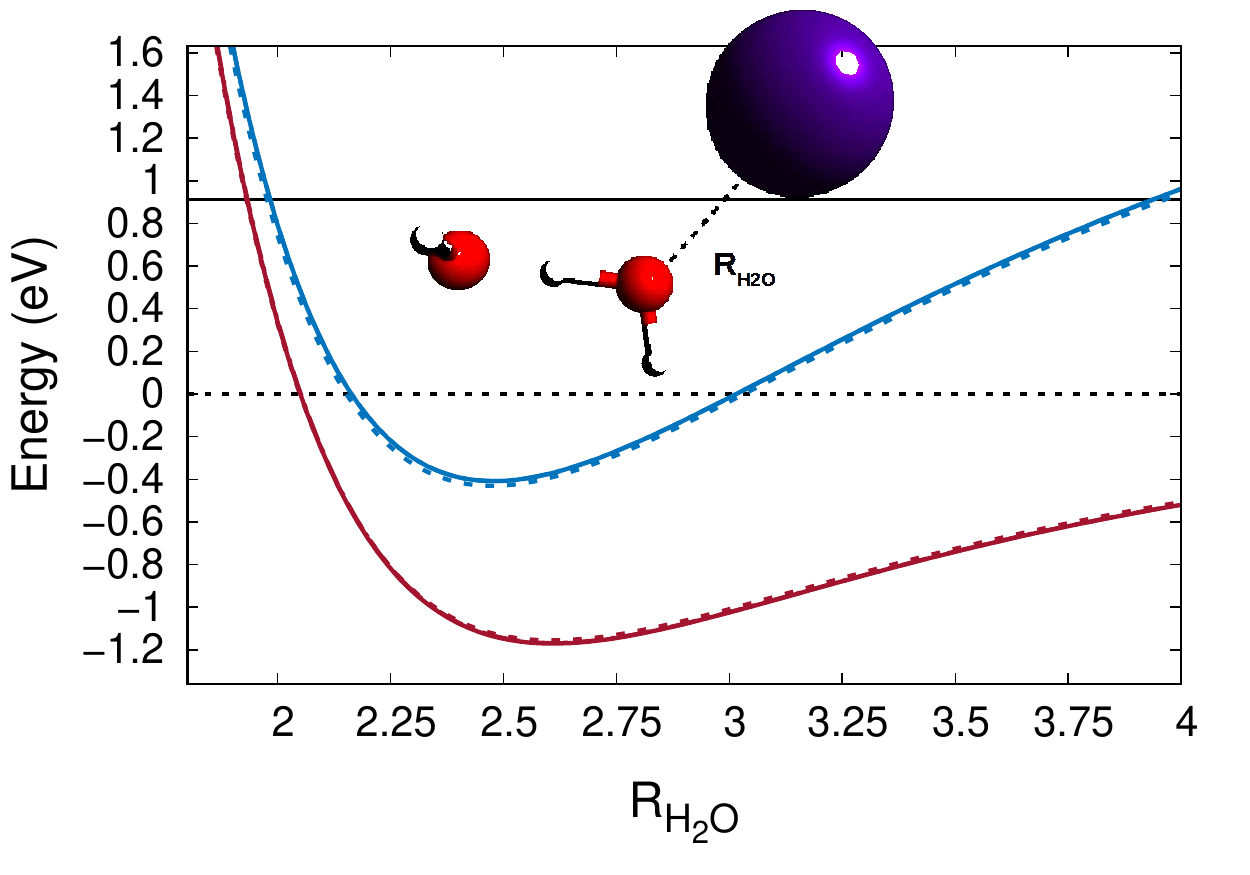}%
\end{subfigure} \\
\begin{subfigure}{0.9\textwidth}
\includegraphics[scale=0.7]{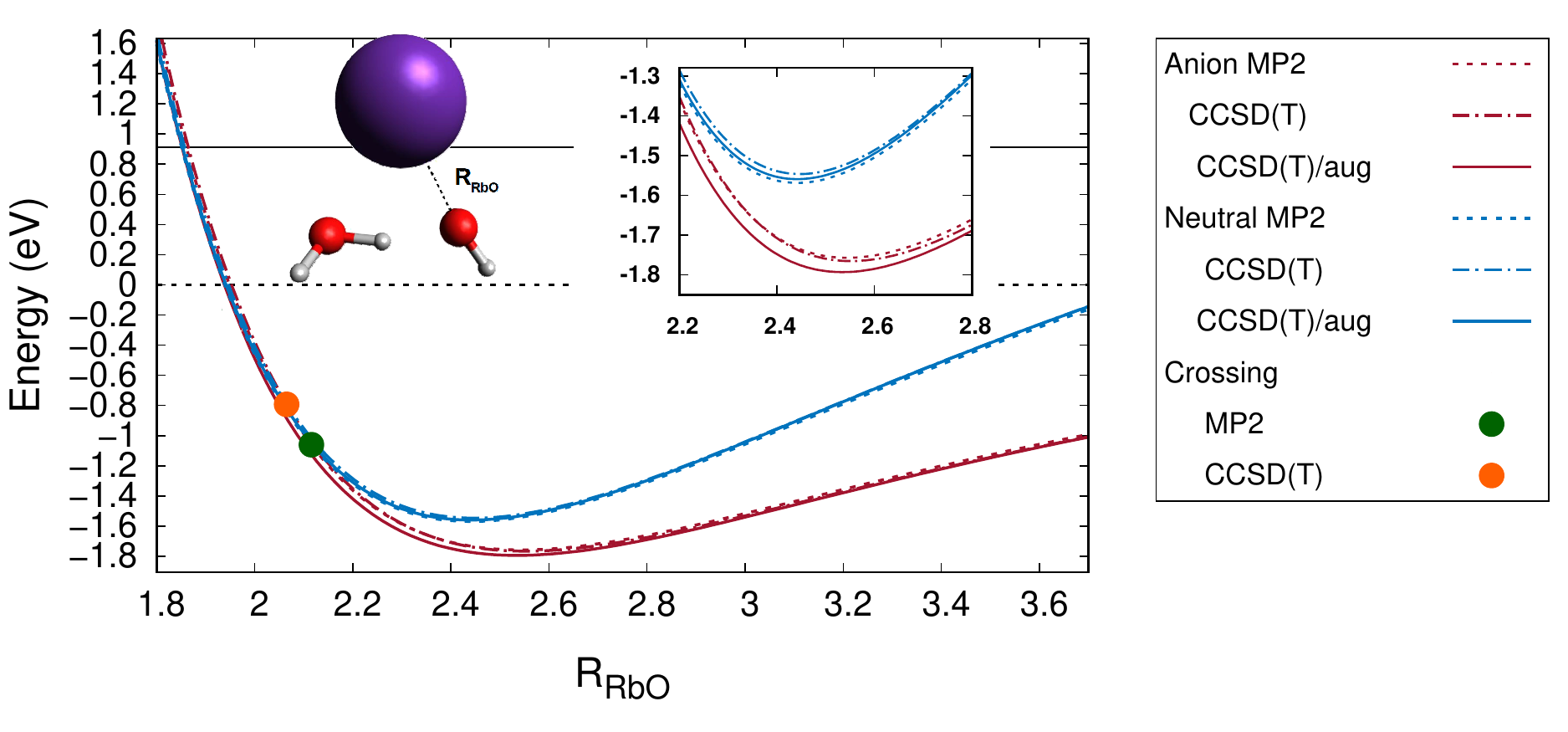}
\end{subfigure}
\caption{MP2 and CCSD(T) neutral (blue) and anion (red) PECs of the Rb-OH(H$_{2}$O)$^{-}$ molecular specie as a function of the Rb-HO distance (upper left), Rb-OH$_{2}$ distance (upper right) and Rb-OH distance (bottom). The coordinates are shown in the figures. The internal coordinates of the cluster have been frozen at their MP2/AVTZ optimized value. The insets on the top right and bottom figures are "zoom" into the potential well. The crossing between the anion and neutral curves are marked by circles for the non augmented case, the crossing lying above 1.6 eV when diffuse functions are used.}
\label{PEC_Rbn1}
\end{figure}

In order to compare our results with the H case, we started by investigating the Rb-OH(H$_{2}$O)$^{-}$ specie. The same coordinates as in the H case have been used for Rb (see Figure \ref{n1_coord}). The R$_{H}$ distance will be labelled R$_{Rb}$ when dealing with Rb. The MP2/AVTZ/MDF$spdfg$ optimized Jacobi coordinates in the frozen cluster approximation can be seen in Figure \ref{Rbcluster} (Rb-n1 case). The structural changes induced by the Rb atom are not as marked as for H. The relaxation of the internal coordinates of the cluster leads to a slight rotation of the OH group, the VDE difference is around 1 meV while the energy difference is even smaller. 
Although the excess electron is predominately bound by the charge-dipole interaction in both cases, the large difference between RbOH$^{-}$ and H$_{2}$O$^{-}$ is not only explained by the larger dipole of RbOH ($\approx$ 6.6 D \cite{Kas2017}) but also by the nature of HOMO that bound the excess electron. In the case of RbOH$^{-}$ the HOMO mainly corresponds to the 5$s$ atomic orbital of Rb with only little contribution from the O and H atomic orbitals \cite{Kas2017}. For the water anion, the HOMO is a mix between 1s$_{H}$ and 2p$_{O}$ atomic orbitals, resulting in a larger deformation of the dipole bound s-like HOMO. \\
Several cuts of the MP2/AVTZ/MDF$spdfg$ PES can be seen in Figure \ref{Map_Rbn1} which corresponds to the same $\beta$ values shown for H (i.e $\beta=$0, 45, 90 and 135 $\degree$, see Figure \ref{Map_Hn1}). The internal coordinates of OH(H$_{2}$O)$^{-}$ have been kept frozen. In addition, we have plotted the energy difference between the neutral and anion PES with the autodetachment region highlighted in dark red (see section \ref{sec_Hn1} for further explanations). The isoline corresponds to the anion's interaction energy with a step of 0.27 eV. The 0 value is shown in dashed line while the energy corresponding to E$_{0}$ (300 K entrance channel energy for vibrational ground state; see section \ref{sec_Hn1} for details) is shown in thick black.  
As can be seen, the AD region is only accessible for some collision angles corresponding to Rb approaching the O atom of the OH group. Although this is also the case for collision involving H, the angular space at which the autodetachment region is reached is considerably smaller for Rb, suggesting smaller detachment cross sections. \\
In order to compare with the result of H in section \ref{sec_H} and probe the effect of the method and basis sets on the interaction energy and AD region, we have calculated the PEC for three different approaches of the Rb. The results can be seen in Figure \ref{PEC_Rbn1}.
The interaction potentials are attractive and rather deep for the three considered configurations. As for the H case, the most favourable collision approaches (with the most accessible AD region) corresponds to the Rb colliding with the O atom of the OH group whereas the Rb-HO collinear configuration corresponds to a maximum. The additional diffuse functions have smaller effects than for H. This can be explained by the more contracted nature of the HOMO due to the larger dipole moment of Rb-OH species (over 6.5 D for RbOH) compared to H-OH (1.85 D for water). Nonetheless, the use of additional diffuse functions stabilizes the anion leading to a higher lying AD region. \\
Unlike the H case, the VDE stays positive when the distance between OH and H$_{2}$O increases. This is not surprising since the RbOH$^{-}$ fragment, obtained upon dissociation of the water group, is stable (RbOH$^{-}$ has a positive VDE). The formation of the RbOH$^{-}$ molecular anion seems therefore to be a possible product of the Rb+OH(H$_{2}$O)$^{-}$ reaction. The energy path diagram for the different product of the Rb+OH(H$_{2}$O)$^{-}$ is shown in Figure \ref{Path_n1}. All fragments have been fully optimized at the MP2/AVTZ/MDF$spdfg$ level of theory and their energy obtained using the CCSD(T) method with the same basis set.
\textsc{
\begin{center}
\begin{figure}[]
\centering
\includegraphics[scale=0.6]{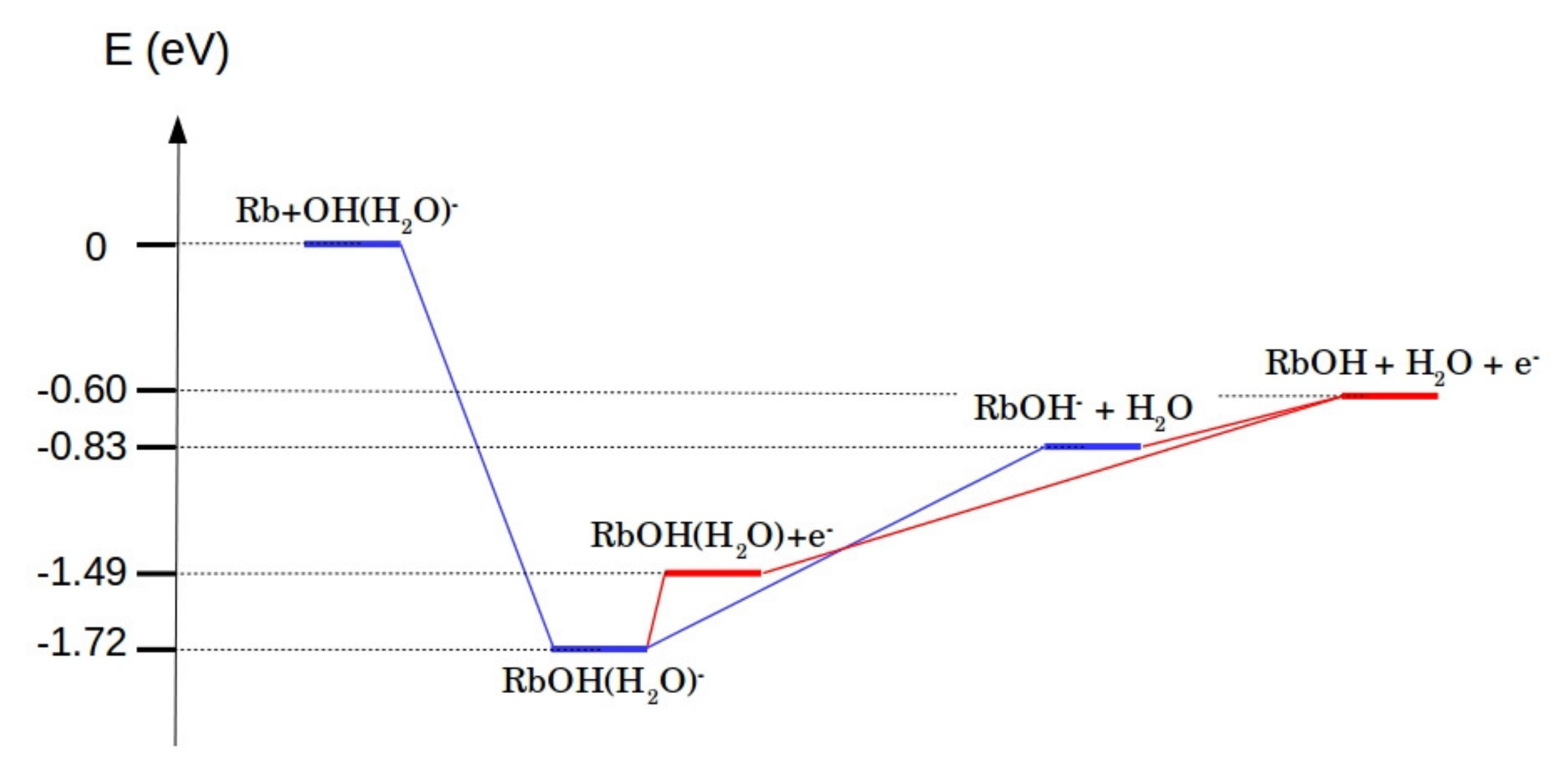}
\caption{Reaction path for the Rb+OH(H$_{2}$O)$^{-}$ reactants. The detachment process are connected with a red lines while the reaction where the excess charge is kept are linked by blue lines.}
\label{Path_n1}
\end{figure}
\end{center}
}   
\noindent All shown products are energetically accessible. However, the production of the direct AED reaction, RbOH(H$_{2}$O), is unlikely since the AD region in only reachable for high collision energy and/or higher internal temperature of OH(H$_{2}$O)$^{-}$. The indirect detachment process involving RbOH$^{-}$ will strongly depend on energy distribution consideration. In particular, the products are expected to be vibrationally hot. In other words, if the ro-vibrational energy of RbOH$^{-}$ is high enough, it may undergo ro-vibrational induced electronic detachment \cite{Simonsl1981}, leading to the RbOH $+$ H$_{2}$O $+$ $e^{-}$ products. According to Figure \ref{Path_n1} the RbOH$^{-}$ molecule will be produced with a maximum excess energy of 0.83 eV (neglecting the kinetic and ro-vibrational energy carried away by the dissociated water molecule). Since the AD region for RbOH$^{-}$ lies at least 1 eV above its minimum \cite{Kas2016}, the probability of (ro-)vibrational induced detachment should be small. We have also investigated the possibility of producing neutral or anionic hydrated Rb: Rb(H$_{2}$O)$^{(-)}$. Although we found this species to be stable (with Rb bound to the O atom), the product channel Rb(H$_{2}$O)$^{-}$+OH and Rb(H$_{2}$O)+OH$^{-}$ are both exothermic by 4.5 eV and 2.4 eV, respectively. Based on this considerations, the production of RbOH$^{-}$ is likely to be the dominant product channel of the Rb+OH(H$_{2}$O)$^{-}$ reaction. \\
 
\FloatBarrier

\subsection{Rb+larger clusters}
\label{sec_Rbn}
     
\begin{figure}[]
\centering
\includegraphics[scale=0.4]{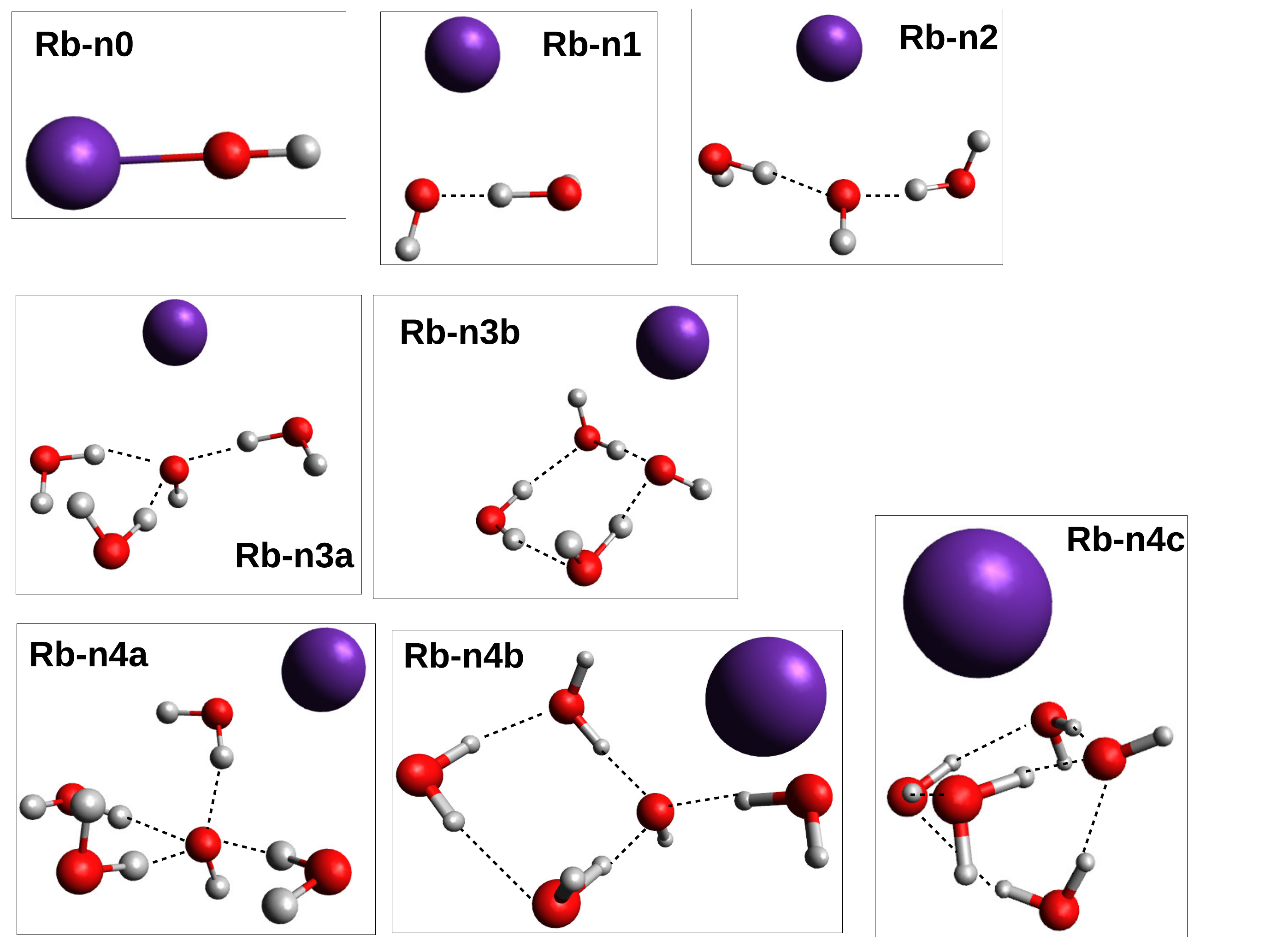}
\caption{Rb approaches geometry leading to the minimum energy for the different hydrated hydroxide anions. Jacobi coordinates of the Rb atom have been optimized at the MP2/AVTZ/MDF$spdfg$ level of theory.}
\label{Rbcluster}
\end{figure}

\begin{figure}
\centering
\begin{subfigure}{1\textwidth}
\centering
\includegraphics[scale=0.8]{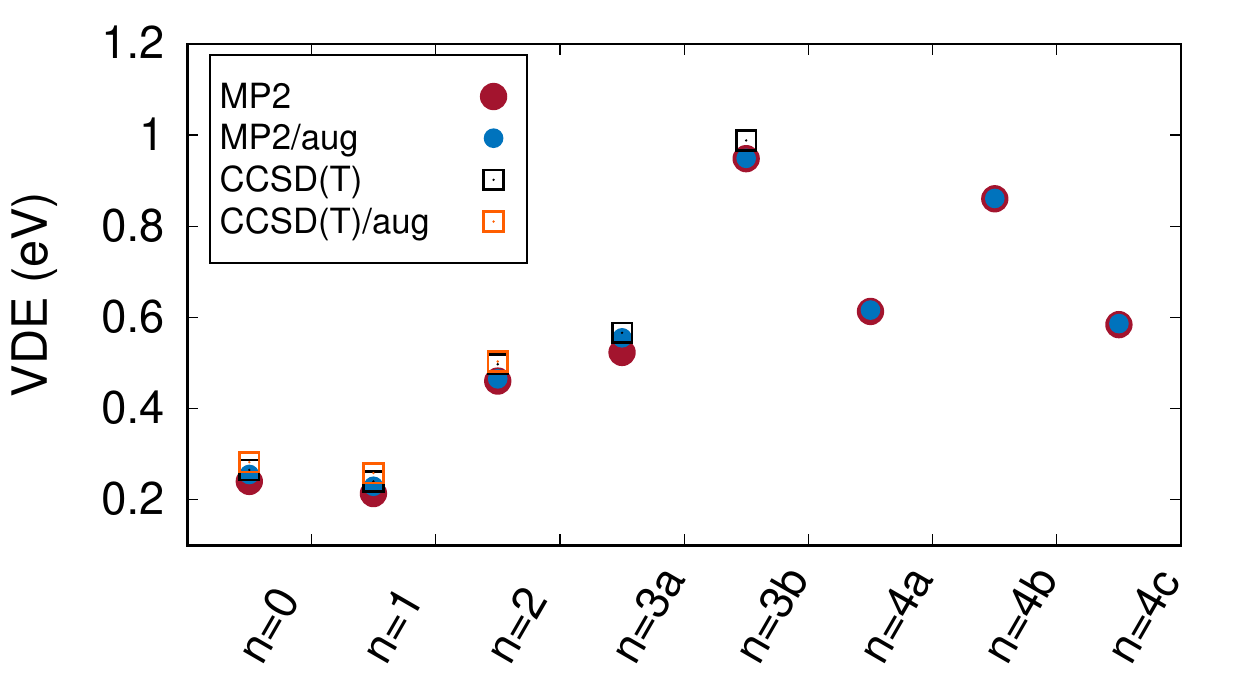}
\end{subfigure}
\centering
\begin{subfigure}{1\textwidth}
\centering
\includegraphics[scale=0.8]{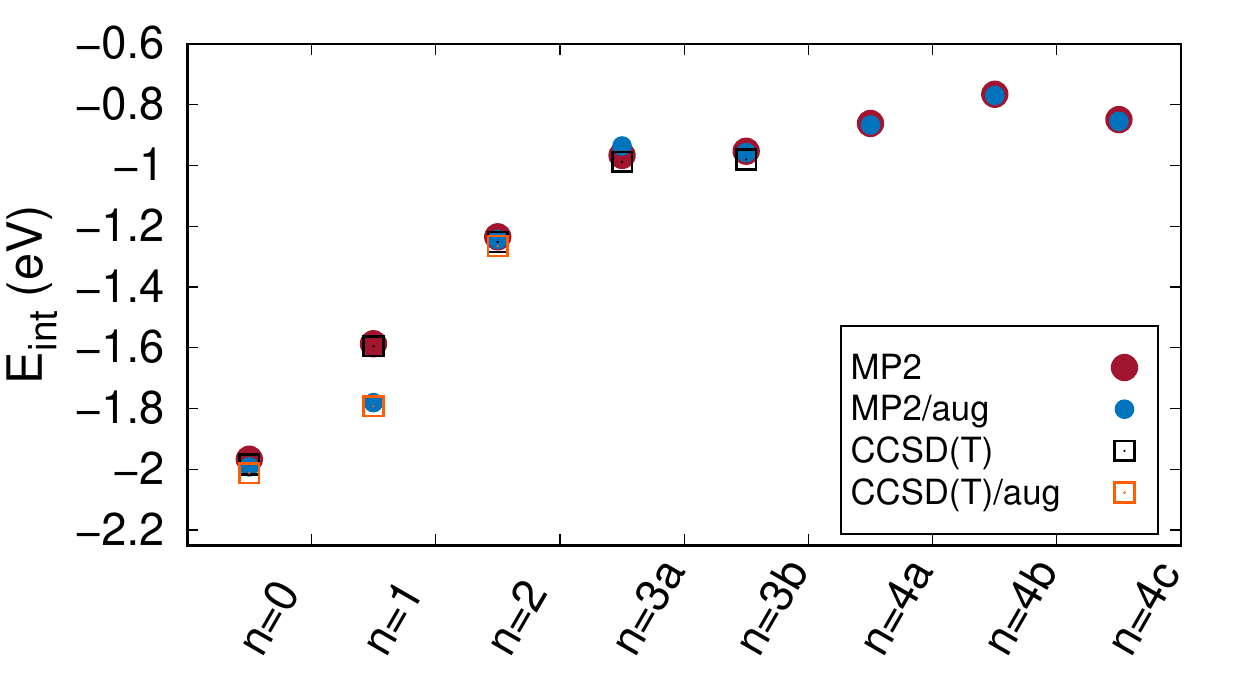}
\end{subfigure}
\caption{MP2 and CCSD(T) vertical detachment energy (top) and interaction energy (bottom) for the Rb-OH(H$_{2}$O)$_{n}^{-}$ system. Calculated for the optimum structures shown in Figure \ref{Rbcluster}. The AVTZ/MDF$spdfg$ basis set has been used.}
\label{EAEint_Rbcluster}
\end{figure}

In order to  study the trend for increasing size of the cluster, we started by performing geometry optimization on the Rb-OH(H$_{2}$O)$_{n}^{-}$ systems. The internal coordinates of the clusters have been kept frozen at their MP2/AVTZ values. The distance R$_{RbO}$ between Rb and the O atom of the hydroxide group as well as the polar and azimuthal angles have been optimized at the MP2/AVTZ/MDF$spdfg$ level of theory. The starting guess geometry corresponds to the Rb atom along the O-H bond from the O side. This procedure is justified by our results on Rb-OH(H$_{2}$O)$^{-}$ in section \ref{sec_Rb_n1} where we have shown that the Rb atom "prefers" approaching the cluster from the O side of the OH group. The results are shown in Figure \ref{Rbcluster}. 
The interaction energy (E$_{int}$) and VDEs at equilibrium geometry have been obtained at the MP2/AVT/MDF$spdfg$+BSSE and CCSD(T)/AVTZ/MDF$spdfg$+BSSE (except for n=4) level of theory. Comparisons are also made with the augmented basis set. The results are shown in Figure \ref{EAEint_Rbcluster}. 
One can see that the VDE of the Rb-OH(H$_{2}$O)$_{n}^{-}$ species also increases with increasing size of the cluster. The same conclusion as for the hydrated hydroxides and water cluster holds here \textit{i.e} the negative charge is stabilized by the water shell. Unlike for the H case, the large differences between some isomers cannot be easily explained and do not follow a clear trend. Similar to the mono hydrated case, additional diffuse functions have only a small effect on both the VDE and E$_{int}$ with the notable exception of the interaction energy of $n1$.  
Before discussing the dissociation energies it is important to note that we have used the frozen monomer approximation, \textit{i.e} the internal coordinate of the OH(H$_{2}$O)$_{n}^{-}$ group has been kept frozen. This will lead to slightly lower bond strengths since we are not on the global minimum. The dissociation energies, shown in Figure \ref{EAEint_Rbcluster}, decrease for increasing size of the cluster, from 2.00 eV for $n=0$ to 0.75 eV for $n=4$. This is most likely due to the decrease in partial charge on the O atoms from the OH group, which leads to a weaker ionic interaction between O and Rb. The ZPE effects have not been taken into account in our dissociation energy calculations. Harmonic frequency calculations on the RbOH(H$_{2}$O)$_{n}^{-}$ smallest species ($n=1$ and 2) lead to a global ZPE roughly 0.1 eV smaller than for the neutral hydrated hydroxide. This difference is likely to be similar for larger clusters. The dissociation energies will thus be around 0.1 eV smaller when taking the ZPE into account. Nevertheless, our calculations provide a qualitative description and unravel the trends in the interaction of Rb with hydrated water clusters. \\
Even though the short-range interaction decreases with increasing size of the cluster, the long-range interaction, important for low energy collisions, should not exhibit such simple behaviour. Indeed, the first classical interaction term is the charge-induced dipole term: $\alpha^{Rb}/2R^{4}$, where $\alpha$ is the polarizabilty of the Rb atom. This term is equal for all Rb-hydrated hydroxide system. However the dipole-induced dipole term $cos\theta \mu^{c} \alpha^{Rb}/R^{5}$ and London dispersion force $\propto \alpha^{Rb}\alpha^{c}/R^{6}$ will depend on the dipole, and polarizability of the cluster. Both properties are expected to depend on their size and will influence the rate constants for elastic and reactive collisions.  \\

\begin{figure}
\centering
\begin{subfigure}{0.48\textwidth}
\centering
\includegraphics[scale=0.6]{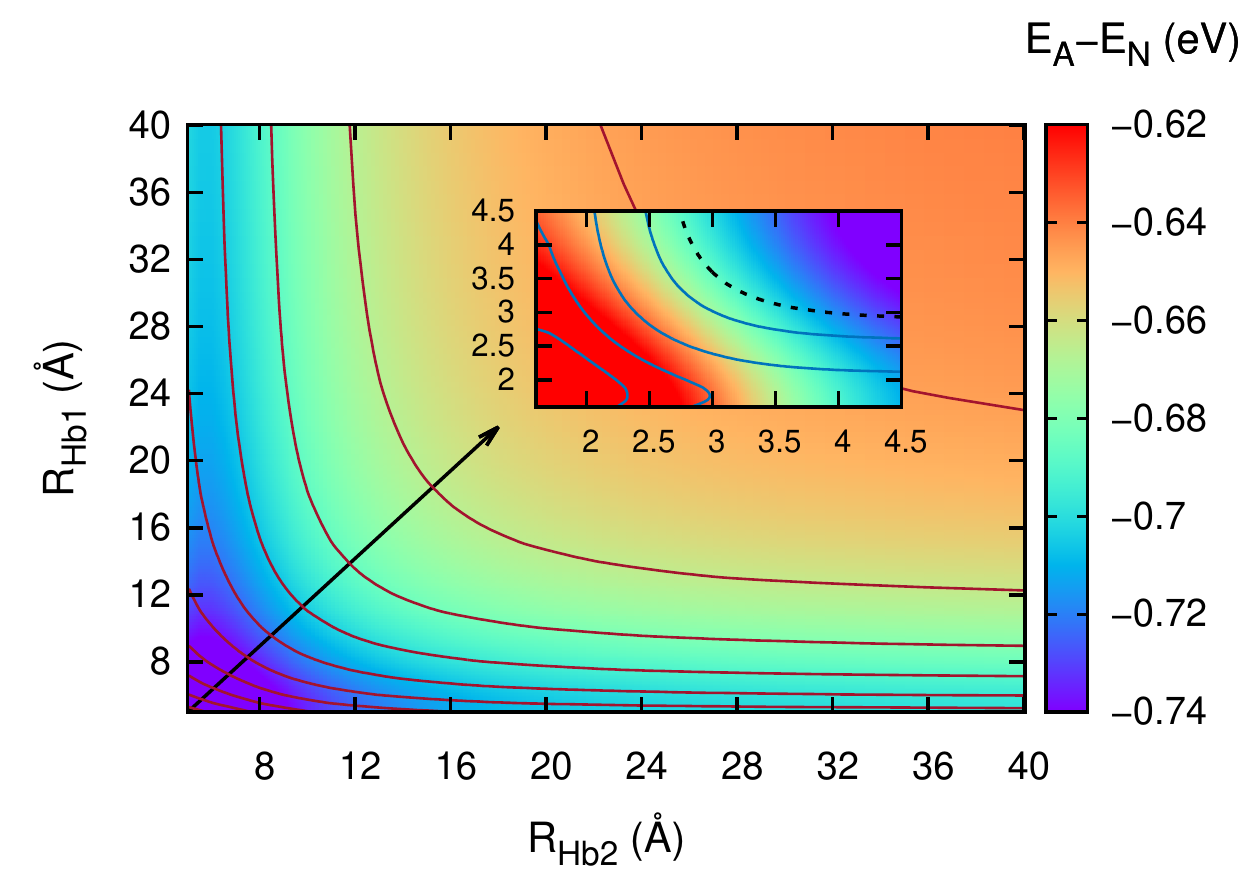}
\end{subfigure}
\centering
\begin{subfigure}{0.48\textwidth}
\centering
\includegraphics[scale=0.35]{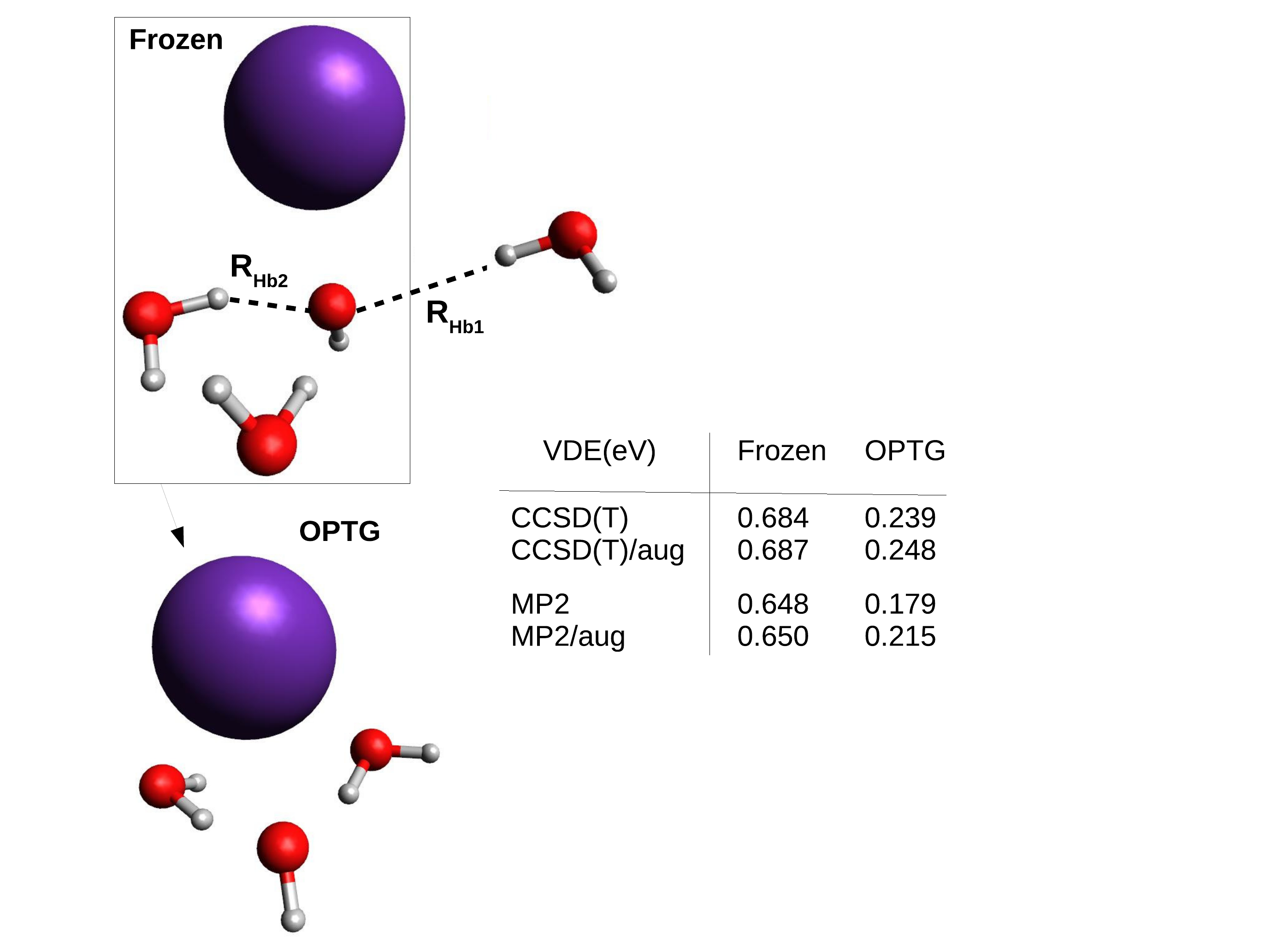}
\end{subfigure}
\caption{Left panel: difference between the anion and neutral MP2/AVTZ/MDF$spdfg$ energy of the Rb-OH(H$_{2}$O)$_{3}^{-}$ specie along the R$_{Hb2}$ and R$_{Hb1}$ coordinates. The isoline corresponds to iso-contour of the anion PES with a step of 0.02 eV starting at 0.01 eV. The energies are given relative to the entrance channel Rb+H(H$_{2}$O)$_{3}^{-}$. The inset shows the results for smaller distances with the isoline starting at -0.7 eV with a step of 0.2 eV. The dashed one corresponds to 0 eV. \\
Right panel: geometry of the investigated specie with the labelling of the R$_{Hb2}$ and R$_{Hb1}$ coordinates (top). MP2/AVTZ/MDF$spdfg$ optimized geometry of the fragment obtained at large R$_{Hb1}$ distance. The calculated CCSD(T) and MP2 VDE are given in the table. The results in parenthesis are obtained at the MP2/AVTZ/MDF$spdfg$+aug level of theory.}
\label{Rbn3_RHb}
\end{figure}

\begin{figure}[]
\centering
\includegraphics[scale=0.5]{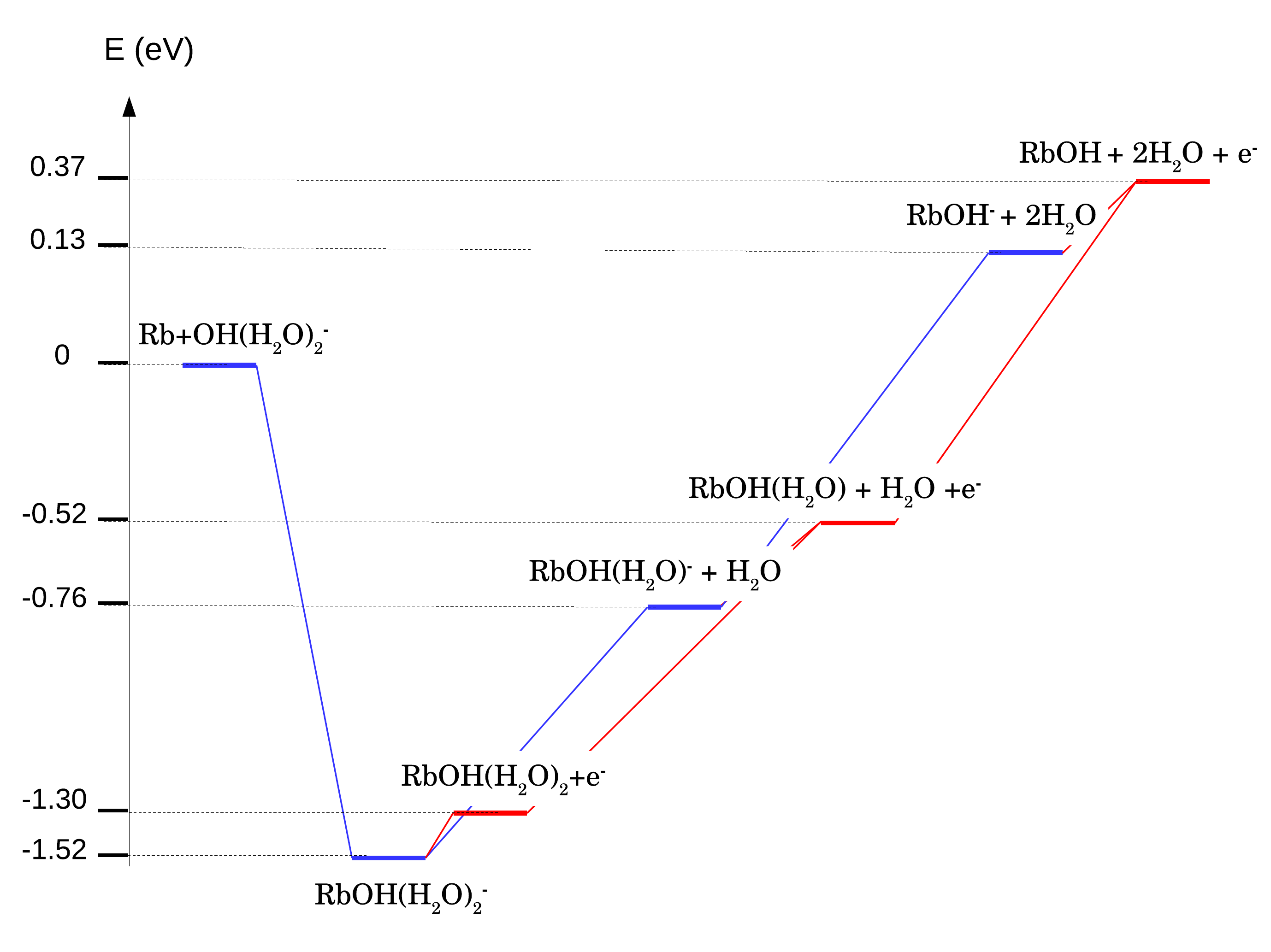}
\caption{Reaction path for the Rb+OH(H$_{2}$O)$_{2}^{-}$ reactants. The detachment process are connected with a red lines while the reaction where the excess charge is kept are linked by blue lines.}
\label{Path_n2}
\end{figure}

\begin{figure}[]
\centering
\includegraphics[scale=0.6]{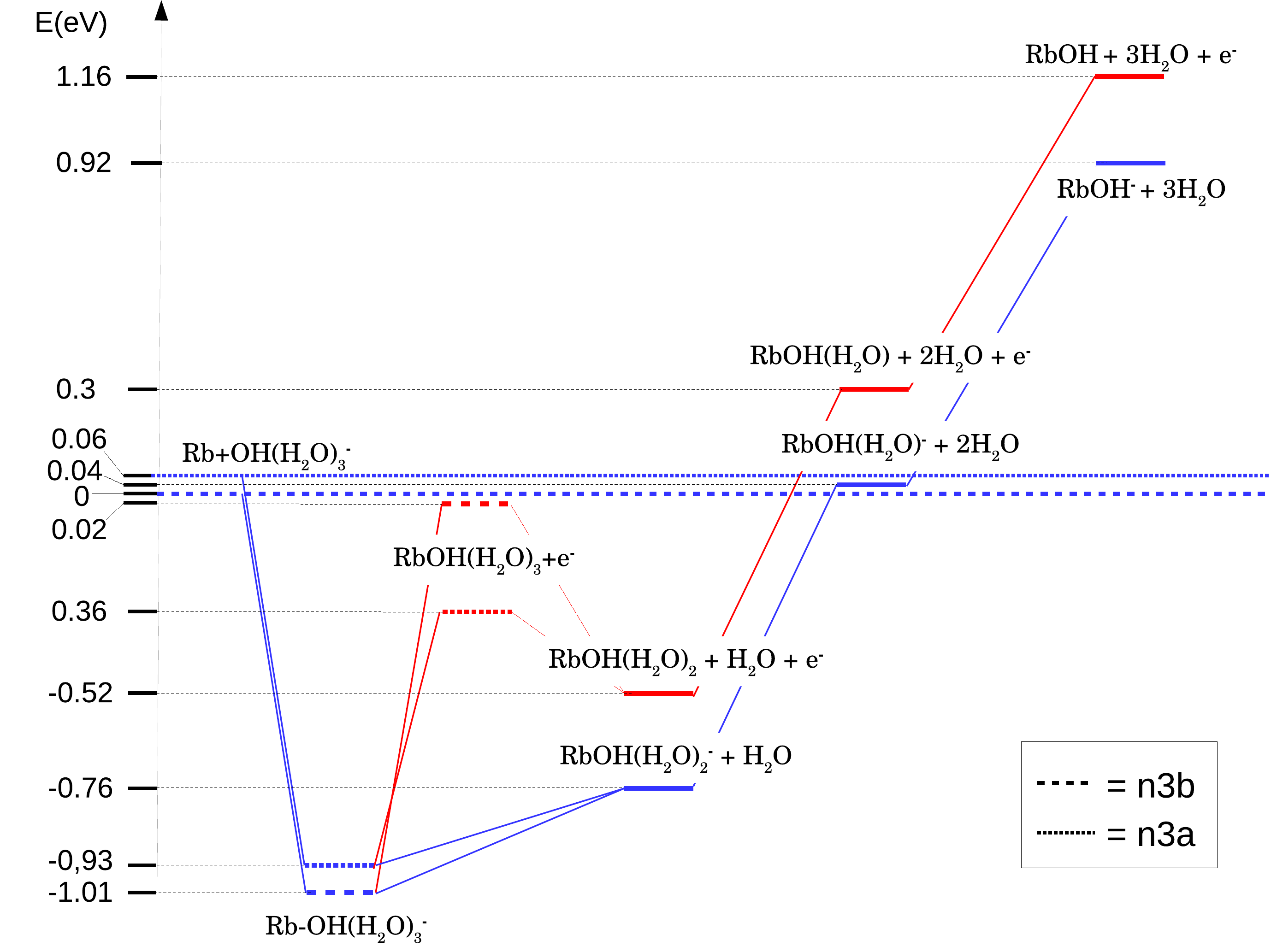}
\caption{Reaction path for the Rb+OH(H$_{2}$O)$_{3}^{-}$ (isomer n3a and n3b) reactants. The detachment process are connected with a red lines while the reaction where the excess charge is kept are linked by blue lines. The zero energy is taken has the energy of the Rb$+$most stable isomer, \textit{i.e} n3b.}
\label{Path_n3}
\end{figure}

\begin{figure}[]
\centering
\begin{subfigure}{0.5\textwidth}
\centering
\includegraphics[width=1\textwidth]{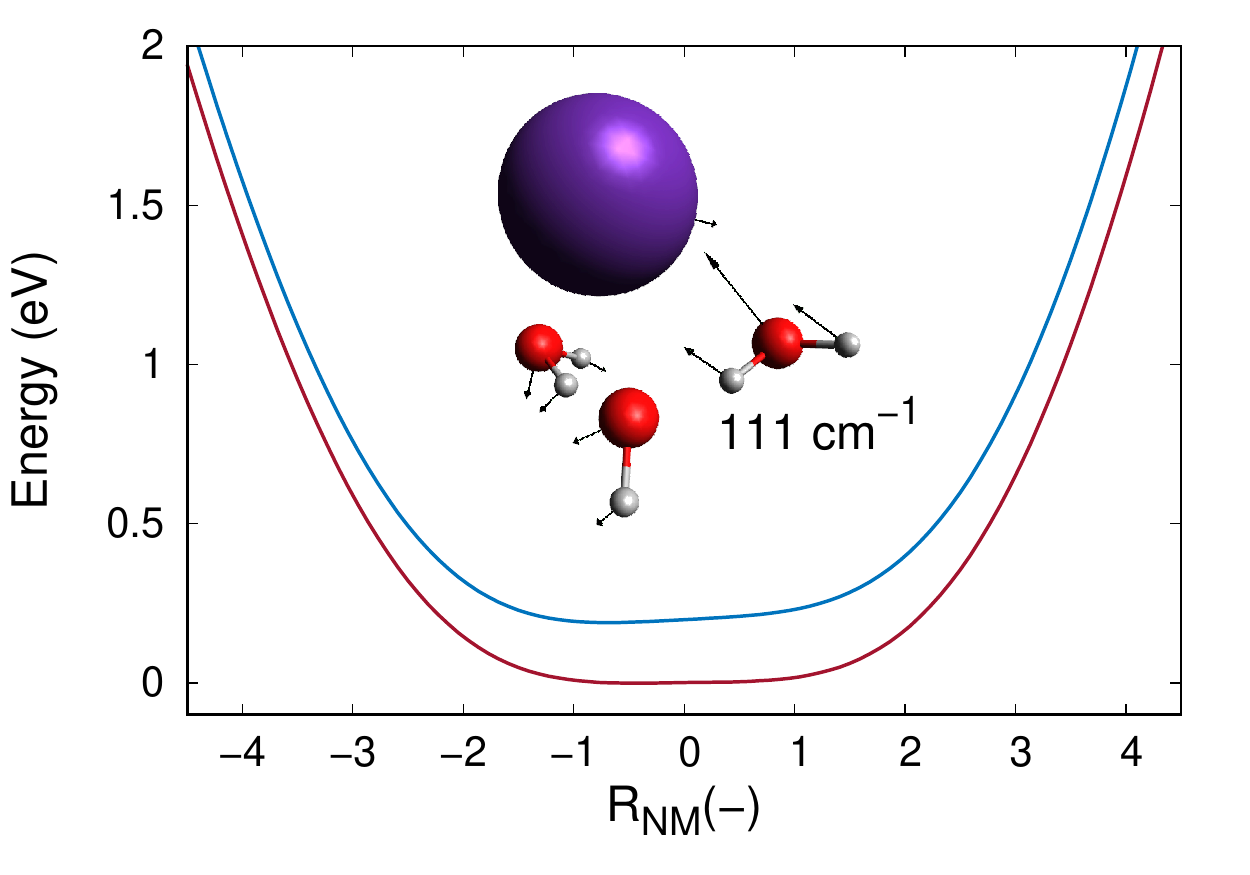}%
\label{NM_111}
\end{subfigure}%
\begin{subfigure}{0.5\textwidth}
\centering
\includegraphics[width=1\textwidth]{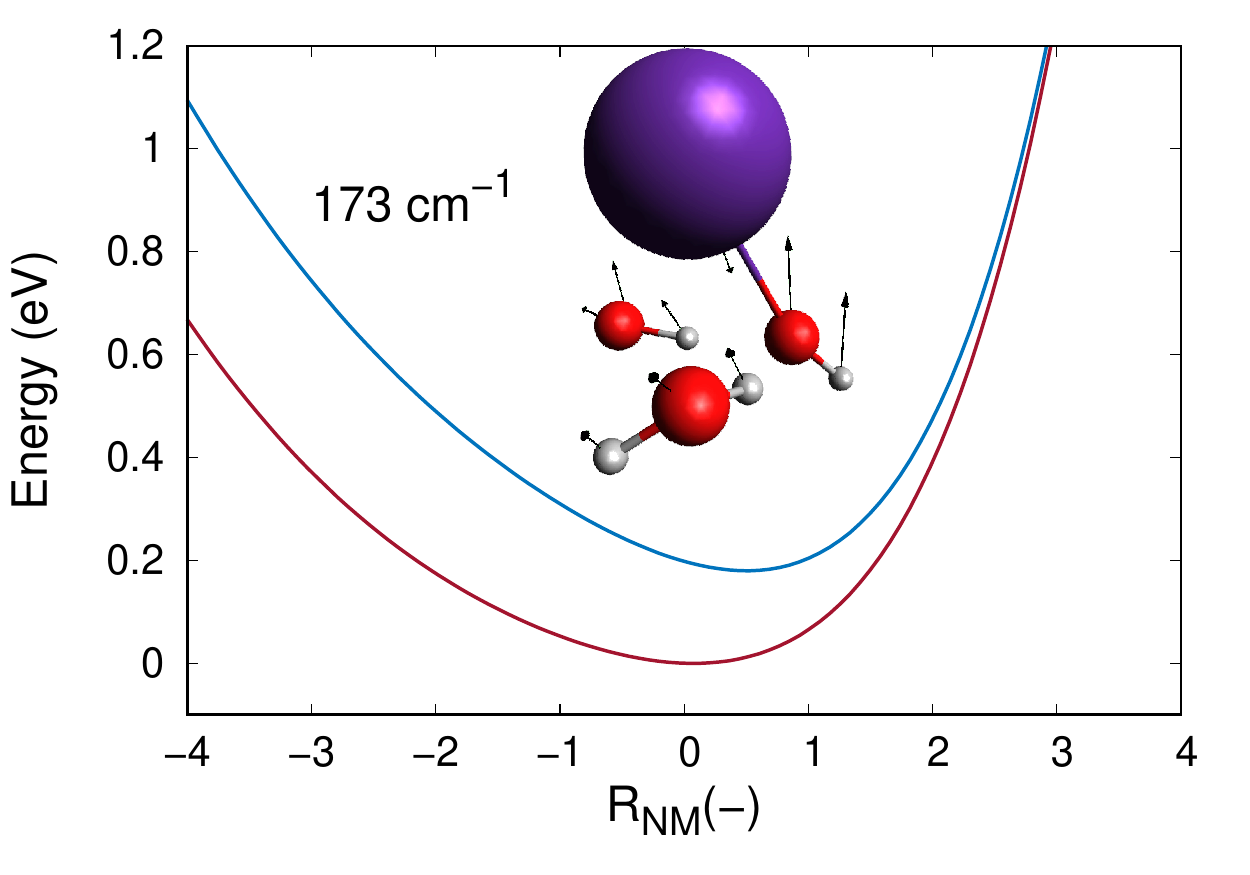}%
\label{NM_173}
\end{subfigure} \\
\begin{subfigure}{0.5\textwidth}
\centering
\includegraphics[width=1\textwidth]{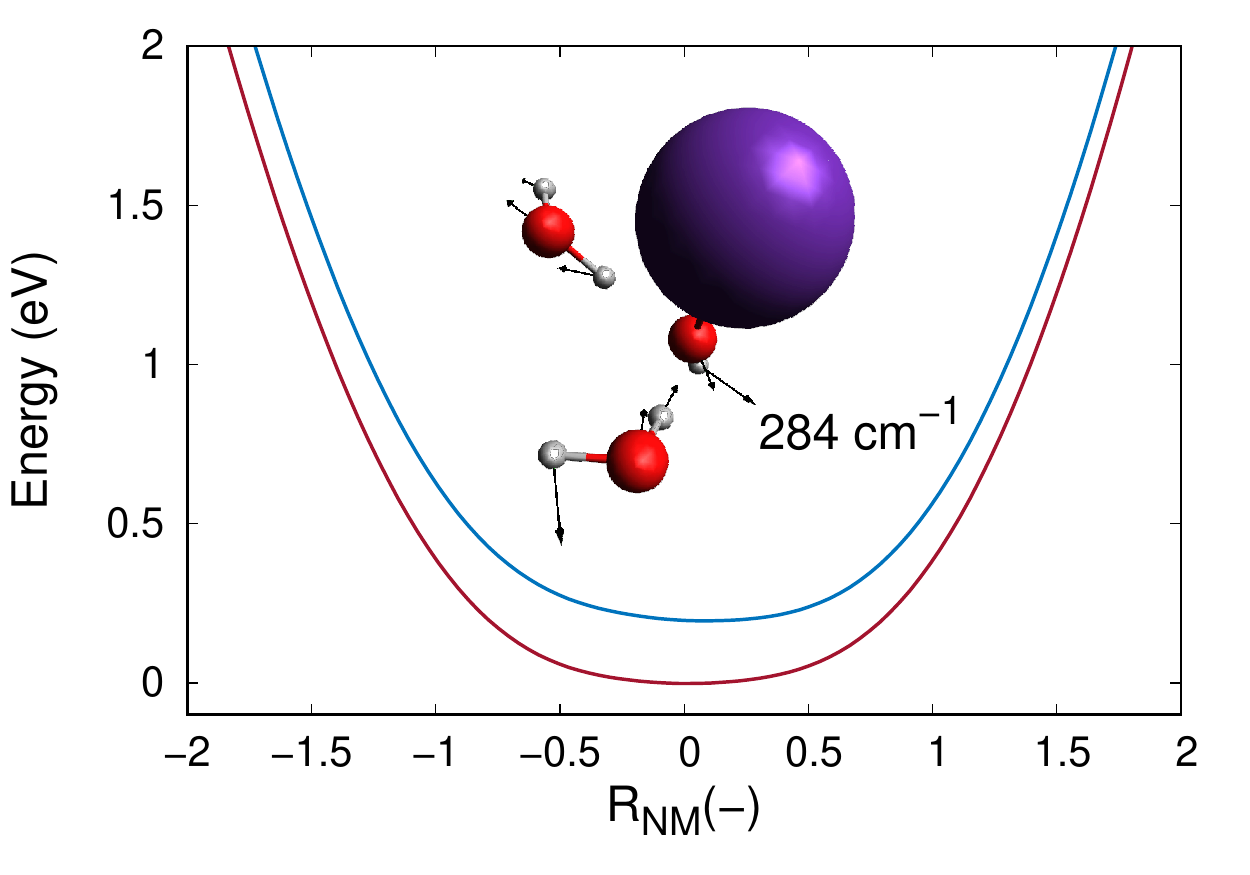}%
\label{NM_284}
\end{subfigure}%
\begin{subfigure}{0.5\textwidth}
\centering
\includegraphics[width=1\textwidth]{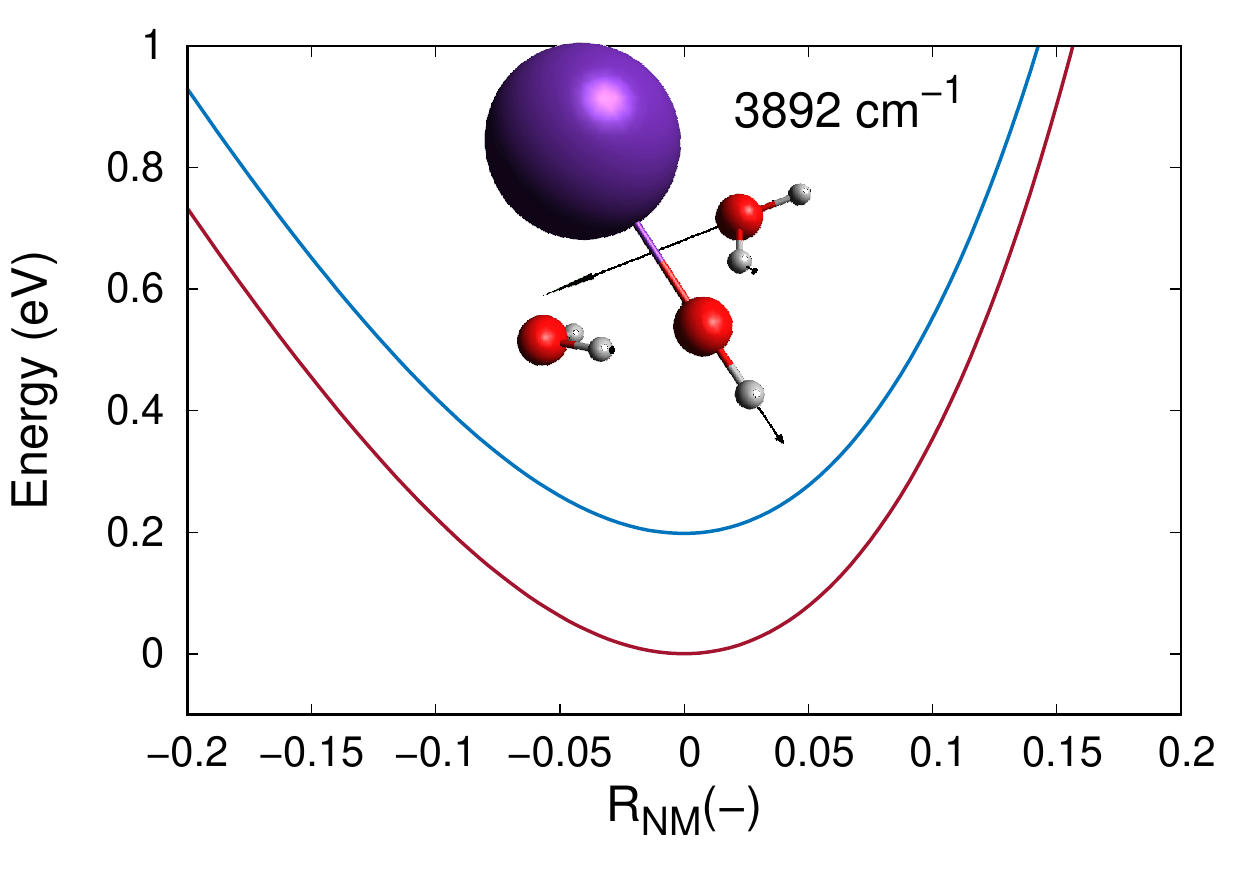}%
\label{NM_3892}
\end{subfigure}%
\caption{MP2/AVTZ/MDF$spdfg$+aug PEC of the neutral (blue) and anion (red} RbOH(H$_{2}$O)$_{2}^{-}$ molecular species along 4 different normal modes coordinates. The x axes corresponds to a fraction of the atomic displacement labeled by black arrows in the molecular drawing. Optimization and frequency calculations were performed at the MP2/AVTZ/MDF$spdf$ level of theory. The normal mode frequency are reported in each plot. 
\label{Rbn2_NM}
\end{figure}

\noindent The direct detachment process, triggered by the crossing between neutral and anion PES, seems to be less effective for collisions involving Rb compared to H. As already pointed out in the previous section, the AD region is already hardly accessible for $n=1$. In order to investigate the trend for increasing size of the cluster, we have calculated the PEC corresponding to the geometry seen in Figure \ref{Rbcluster} as a function of $R_{RbO}$, the distance between Rb and the O atom of the OH group. The curves have been obtained at the MP2/AVTZ/MDF$spdfg$+aug+BSSE level of theory with the internal coordinates of the clusters kept frozen. The coloured horizontal lines corresponds to the energy in the entrance channel E$_{0}$ (see section \ref{sec_Hn1} for further explanation).
\begin{figure}[]
\centering
\includegraphics[scale=0.8]{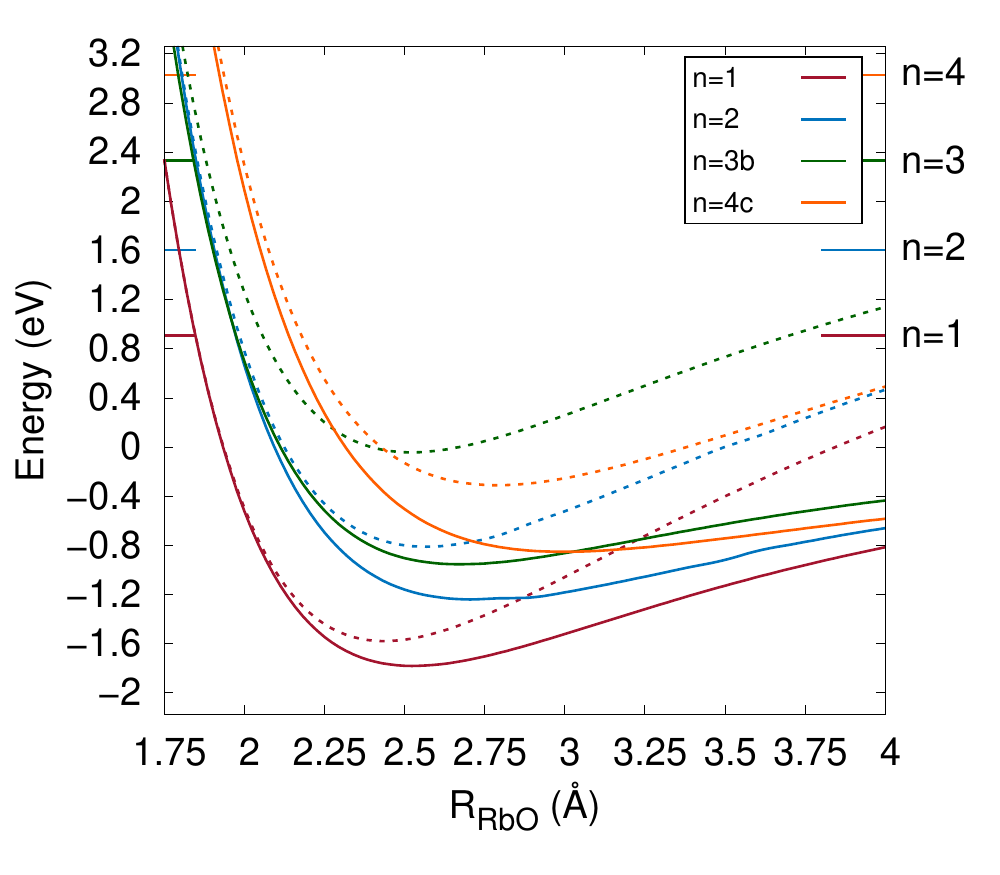}
\caption{MP2/AVTZ/MDF$spdfg$+aug energy of the Rb-OH(H$_{2}$O)$_{n}^{-}$ systems as a function of R$_{RbO}$ in their preferred approach configurations (see Figure \ref{Rbcluster}. The coloured horizontal lines corresponds to E$_{0}$, the energy at the entrance channel (see text for further explanations).}
\label{Rbn_cros}
\end{figure}
The results show a clear trend, with the crossing point, hence the AD region, lying higher in energy for increasing size of the cluster. This can related to the trend in VDE (see Figure \ref{EAEint_Rbcluster}). Our results suggest very low direct detachment probability, especially considering that the investigated collisional approach leads to the lowest lying autodetachment region. The results for the different isomers are given in supplementary material. Unlike for the H case, no potential barriers are found along the R$_{Rb}$ coordinate. Furthermore, the influence of the isomer on the interaction energy and the AD region is less clear than in the case of H. This can be explained by Rb being less sensitive to its chemical environment as shown in Figure \ref{Map_Rbn1} for Rb-OH(H$_{2}$O)$^{-}$ where the different interaction energies range from -0.54 eV to -1.6 eV. \\          
Analogously to the H case, we have considered the indirect detachment mechanism. The anion-neutral energy difference of the Rb-OH(H$_{2}$O)$_{3}^{-}$ was scanned along 2 water dissociations coordinates: R$_{Hb2}$ and R$_{Hb2}$. The calculation was performed at the MP2/AVTZ/MDF$spdfg$ level of theory with all other coordinates frozen. The VDE of the fragment, obtained upon dissociation of one water molecule, has been calculated at various levels of theory with and without global optimization. The results are depicted in Figure \ref{Rbn3_RHb}.      
The anion PES lies below the neutral for the entire scanned nuclear coordinates, \textit{i.e} E$_{A}-$E$_{N}$ stays negative. This suggests the formation of stable negatively charged species, which is confirmed by the calculated VDE of the fragment obtained for R$_{Hb1} \rightarrow \infty$ (see right table in Figure \ref{Rbn3_RHb}). \\ 
In addition, all dissociation products lie above the energy of the entrance channel, in other words the reactions are endothermic. In fact, as already mentioned above, the Rb-OH bond is weaker than the H-OH bond. The amount of energy produced by the bond formation is thus not enough to break all water bonds. Figure \ref{Path_n2} and \ref{Path_n3} shows the energetic diagram for the dissociation and detachment channels of the dihydrated Rb+OH(H$_{2}$O)$_{2}^{-}$ and trihydrated Rb+OH(H$_{2}$O)$_{3}^{-}$, respectively. All energies have been obtained with the CCSD(T) method and AVTZ/MDF$spdfg$ basis. The energies have been calculated on the MP2/AVTZ/MDF$spdfg$ global optimized geometry for all species except the Rb-OH(H$_{2}$O)$_{3}^{-}$ complexes for which convergences were not reached. We therefore took the energy at the frozen cluster geometry for the latter (see Figure \ref{Rbcluster}). The exit channels for hydrated Rb products are not shown since they all lie higher in energy, the Rb(H$_{2}$O) bond being weaker than the RbOH bond.     
As predicted on the basis of the Rb-OH bond strength, the total dissociation products are endothermic. Furthermore, in the trihydrated case Rb+OH(H$_{2}$O)$_{3}^{-}$ ($n3a$ or $n3b$), the exit channel corresponding to RbOH(H$_{2}$O)${-}$+2H$_{2}$O is endothermic for collision with isomer $n3b$ while exothermic for isomer $n3a$, hence suggesting isomer dependency. The possibility to produce RbOH(H$_{2}$O)$_{l}^{-}$  species (with $l<n$) will depend on their stability against vibrational induced detachment (VID). Unlike water cluster anions, the latter is predicted to be very small for the following reasons: i) all RbOH(H$_{2}$O)$_{l}^{-}$ cluster have larger VDE compare to water cluster HOH(H$_{2}$O)$_{l}^{-}$, ranging from 0.3 eV to 1 eV, ii) the VID becomes important when the vibrational motion spans part of the AD region, \textit{i.e} an excited vibrational mode brings the molecule into geometries where E$_{anion} > $E$_{neutral}$. Our calculation shows that crossing between anion and neutral PEC relative to the R$_{RbO}$ distance lies very high in energy. Hence, vibrational excitation of the normal mode corresponding to Rb-O stretching should not lead to significant detachment probability, iii) Along all nuclear coordinates investigated in the present study, the PECs of the neutral are steeper than the anion's, \textit{i.e} the VDE increases with structural deformation. This assumption is based on the difference in VDE between frozen and relaxed geometries. The global minimum exhibits smaller VDE. 
In order to verify the latter assumption, we have scanned the neutral and anionic PEC of the optimized RbOH(H$_{2}$O)$_{2}^{-}$ along the coordinates of several vibrational modes. The PEC have been obtained at the MP2/AVTZ/MDF$spdfg$+aug level of theory and are shown in Figure \ref{Rbn2_NM}. The molecular drawings insets shows the atomic displacement within the given normal mode.
In absence of such crossings, the detachment can only be triggered by non-adiabatic couplings. This couplings are larger for very diffuse anionic states for which the associated wave function undergoes rapid changes upon nuclear deformation \cite{Simons2005}, when the anionic and neutral PES are close and parallel and for light atoms. In this regard, it is particularly interesting to point out the difference between hydrated Rb hydroxide anions and water cluster anions. Indeed, the latter are more subject to vibrationaly induced detachment (see for example the water dimer anion \cite{Kim1999}) which can be explained by their lower VDE, implying a smaller energy difference between anionic and neutral PES, but also by molecular orbital considerations. While the HOMO of RbOH(H$_{2}$O)$_{l}^{-}$ compounds is mainly formed by the $5s$ atomic orbital of Rb, it mixes with $2p$ orbitals of O in water cluster anions, leading to larger sensitivity upon nuclear deformations. In addition, the large mass of Rb leads to a smaller probability of vibrationaly induced detachment for hydrated Rb hydroxide anions compared to water cluster anions.  

\FloatBarrier

\section{Conclusions}

We have investigated the interaction energies and autodetachment region of the collisional complexes M+OH(H$_{2}$O)$_{n}^{-}$ for M=H and Rb, and $n=1$ to 4. The main reactive channel at low collision energy is the associative electronic detachment (AED) reaction involving water loss where the only energetically accessible products are MOH bound species. The AED reaction may occur through a direct or indirect mechanism depending on where the detachment of the excess electron takes place. We have performed geometry optimization, calculated interaction energies, vertical detachment energies and cuts of potential energy hypersurfaces for which the effect of diffuse functions and methods have been investigated. Based on our calculations we can make the following conclusions:
\begin{itemize}
\setlength\itemsep{1em}
\item The reactive center is the OH$^{-}$ group where M preferentially bonds to the O atom while the M-H configuration corresponds to a maximum. For M=Rb, the interaction potential is rather deep, ranging from -2 eV to -0.5 eV while the anisotropy is more marked for M=H where it ranges from -3.5 eV to 0.6 eV.  
\item The indirect mechanism is likely to be the dominant pathway for the AED reaction. In particular, a direct detachment process will only be possible for the smallest clusters, \textit{i.e} $n1$ and $n2$, and/or when including excited vibrational states and higher collision energies.   
\item The indirect mechanism leads to the production of vibrationally hot anions and neutrals MOH(H$_{2}$O)$_{l}$ species. The value of $l$ will mainly depend on energetic considerations. For M=H, the total water loss exit channel $(n+1)$H$_{2}$O is open while the weaker Rb-OH bond only allows for partial water loss. Therefore, $l>0$ species are expected when considering $n>2$ hydrated hydroxide clusters colliding with Rb. 
\item The likelihood of obtaining RbOH(H$_{2}$O)$_{l}^{-}$ products will depend on their stability against vibrational induced detachment. Although our results suggest these species to be stable enough, additional calculations need to be performed. In particular, since no crossing between anionic and neutral PES are seen, the detachment will only be triggered by non-adiabatic couplings. 
\item Considering that the AED channel for M=H leads to total water loss and that H$_{2}$O$^{-}$ is unstable against detachment, the production of water cluster anions is unlikely. However, one may observe such species when considering collisions with larger hydrated hydroxide anions. 
\item We found some indications of isomer-dependent reactivity.     
\end{itemize} 
Molecular dynamics methodologies may be suited to explore the wide range of possible reactive pathways and extract rate constants \cite{Tachikawa2014}. However, in order to correctly describe the delocalized nature of the H bond, methods accounting for quantum dynamics effect such as path integral approaches should be preferred. In addition, the detachment process must be taken into account. The simplest way would be to add an absorbing potential at the crossing point between anion and neutral PES (entrance of the AD region) and assume slow ejected electron to model energy distribution of the AED products. \\

\paragraph{Implicationss for hybrid trap experiment}

Our results demonstrate the rich possibilities of studying hydrated hydroxide in the context of hybrid trap experiments. First of all, the hydrated hydroxide cluster themselves, without considering interaction with the ultracold atomic cloud are interesting species. Indeed, their VDE could be measured and the isomer dependency highlighted. \\
\noindent When the trapped ions are immersed into the atomic cloud, elastic, inelastic and reactive collisions may take place. The easiest way to study the latter is through loss measurements \cite{Deiglmayr2012}. Theoretical investigation of the AED reaction between Rb and OH$^{-}$ have shown that the reaction is rather small \cite{Kas2016} whereas experimental measurements are ongoing. As mentioned in the general conclusion above, the reaction between hydrated hydroxide anions and Rb should predominantly lead to RbOH(H$_{2}$O)$_{l}^{-}$ species accompanied by water loss, where $l$ depends on the initial cluster size. If a MOT laser is present, these hydrated Rb hydroxide anions will be photodetached since their VDE range from 0.2 eV to 1 eV, thus below the energy of the MOT laser ( around 1.56 eV for Rb). In this situation, loss rates involving ground state Rb will likely to be larger than for the non hydrated case (OH$^{-}$). The loss rate should then decrease with increasing $n$ due to steric effects, as has been shown for H \cite{Howard1975}. If an other trapping method such as a dipole trap is used for the Rb cloud, observation of the RbOH(H$_{2}$O)$_{l}^{-}$ molecular species may be feasible if they stay trapped long enough to be detected.   \\
\noindent The internal temperature of the anions may be measured by means of threshold photodetachment spectroscopy. However, since the rotational constants of the cluster are much smaller than for OH$^{-}$ (see Table \ref{cluster_prop}), the rotational levels will be very close in energy, making it difficult to resolve the different transitions. In addition to rotation, the clusters also undergo vibrational motion. Some vibrational modes have sufficiently low frequencies to support significant population on the excited states at room temperature. Harmonic frequencies of normal modes are available throughout the literature \cite{Lee2004}. In particular, an extensive study of the OH(H$_{2}$O)$^{-}$ cluster has been carried out using high level quantum dynamics calculations \cite{Pelaez2014} from which accurate anharmonic frequencies have been obtained. Threshold photodetachment may therefore be used to probe the vibrational quenching of the clusters by collision with a buffer gas (such as Rb), as it has already been shown for the OH(H$_{2}$O)$^{-}$-He system \cite{Otto2013}. A recently developed method \cite{Loreau2018} for calculating elastic and inelastic collisions, suited for strongly interacting systems at low temperature may be used for the Rb+OH(H$_{2}$O)$^{-}$ system. More drastic approximations will however be needed to tackle larger clusters. 
In addition, for $n=2$, $n=3$ and $n=4$ (and possibly larger clusters), it should be possible to study specific isomers by using photodetachment laser with appropriate wavelength.
\noindent Finally, based on similarity of the interaction potential and VDE \cite{Kas2017,Tomza2017}, our conclusions are expected hold for different alkali and alkali-earth atoms. \\

\paragraph{Implications for astrochemistry}

Expanding cometary atmospheres are ideal environment for the production of water clusters, water being the most abundant specie. Based on the fact that OH$^{-}$ has been detected in cometary atmospheres \cite{McCarthy2006} and that OH(H$_{2}$O)$_{n}^{-}$ clusters are easily produced in such pressure and temperature conditions \cite{Bykov2013}, one may reasonably suggest the presence of hydrated hydroxide anions in coma of comets. This will have two main impacts on chemical models. First of all, the hydration of OH$^{-}$ should increase their lifetime by preventing collisions with other atoms or molecules. Especially, AED reactions with H atoms are significantly decreased \cite{Howard1975}. Secondly, the hydrated OH$^{-}$ also, perhaps paradoxically, allows for the formation of anionic and neutral MOH bound species through associative reactions followed by water loss and possible electron detachment. The inclusion of water clusters species will surely affects the chemistry of anions and are certainly worth a deeper theoretical and experimental investigation.

\section{\label{sec:akn}Acknowledgment}
The Fonds National de la Recherche Scientifique de Belgique (FRS-FNRS) is greatly acknowledged for financial support (FRIA grant and IISN 4.4504.10 project). 
Computational resources have been provided by the Shared ICT Services Centre, Universit\'e libre de Bruxelles, and by the Consortium des \'Equipements de Calcul Intensif (C\'ECI), funded by the Fonds de la Recherche Scientifique de Belgique (F.R.S.-FNRS) under Grant No. 2.5020.11.

\bibliographystyle{unsrt}
\bibliography{Paper_cluster.bib}

\section{Supplementary material}

\subsection{Hydrated hydroxide}

The VDE of the anion, the dipole moment of the neutral core, the partial charge on the O atom of the hydroxide group and the rotational constant of the OH(H$_{2}$O)$_{n}^{-}$ clusters are reported in Table \ref{cluster_prop} along with the value obtained by Masamura \cite{Masamura2002a} and S. Guha et al. \cite{Guha2014} using the MP2/AVDZ and DFT methods, respectively, and the CCSD(T) results from Morita \textit{et al.} \cite{MORITA2016}. 
\begin{table}[h!]
\centering
\resizebox{\columnwidth}{!}{
\begin{tabular}{c|ccccccc|c|c|ccc}
  & \multicolumn{7}{c|}{VDE (eV)} & $\mu_{e}$(Debye) & $q_{O}$ & \multicolumn{3}{c}{$B_{e}$ (cm$^{-1}$)} \\
   & HF & \multicolumn{4}{c}{MP2} & \multicolumn{2}{c}{CCSD(T)} & MP2                  & HF         & x      & y      & z  \\
 n & AVTZ & AVTZ & AVQZ & Ref.\cite{Masamura2002a} & Ref.\cite{Guha2014} & AVTZ & Ref.\cite{Morita2014} & AVTZ & & & & \\
\hline \\
0  &-0.128 & 2.057 & 2.119 & -    & -    & 1.744 & 1.63 & 1.09  & -1.091  & -      & -      & 18.997 \\                                      
1  & 1.997 & 3.918 & 3.918 & 3.62 & 3.27 & 3.460 & 3.31 & 1.956 & -0.856  & 0.312  & 0.3099 & 10.267 \\
2  & 2.820 & 4.771 & 4.771 & 4.56 & 4.32 & 4.337 & 4.21 & 0.349 & -0.781  & 0.0793 & 0.0835 & 1.2295 \\
3a & 3.563 & 5.463 & 5.535 & 5.18 & 4.98 & 5.109 & 5.07 & 1.771 & -0.760  & 0.0516 & 0.0645 & 0.2247 \\
3b & 3.481 & 5.327 & 5.393 & 5.15 & -    & 4.953 & -    & 1.664 & -0.716  & 0.0644 & 0.1163 & 0.1393 \\
4a & 4.115 & 6.025 & -     & 5.81 & 5.51 & 5.674 & 5.53 & 2.488 & -0.812  & 0.0390 & 0.0572 & 0.0924 \\
4b & 4.147 & 6.017 & -     & 5.69 & -    & 5.655 &      & 1.645 & -0.755  & 0.0365 & 0.0488 & 0.1049 \\
4c & 3.963 & 5.836 & -     & 5.76 & -    & 5.474 &      & 1.031 & -0.716  & 0.0690 & 0.0640 & 0.0860  \\
\end{tabular}
}
\caption{VDE, electric dipole moment, partial charge on the O atom of the OH group and rotational constants of the MP2/AVTZ optimized hydrated hydroxide clusters OH(H$_{2}$O)$_{n}^{-}$ for n=0 to n=4. Note that the dipole moments, $\mu_{e}$, have been obtained using the center of masses as origin for the coordinates.}
\label{cluster_prop}
\end{table}
The dipole moment varies greatly between the structures. The lowest value is found for OH(H$_{2}$O)$_{2}^{-}$, which can be explained by its highly symmetric structure with the two water groups situated on both sides of the OH group. As already pointed in the main text (see section \ref{sec_cluster}), as opposite to water cluster anion where the excess electron is bound by dipole interactions, it is mainly the electronic correlation that binds the excess electron in the hydrated hydroxide anions. They can thus be categorized as valence-bound anions. This explains the large difference between HF and MP2/CCSD(T) results, the relative insensitivity to additional diffuse functions (not shown in the Table) and it also explains why the dipole moments and the VDEs are not correlated. As already discussed, the partial charge on the O atom of the hydroxide group decreases with increasing size of the cluster, leading to the stabilization of the excess negative charge. The calculated VDE for OH$^{-}$ is quite far from the experimental value (1.82 eV \cite{Smith1997}). However, the inclusion of core correlation and the use of larger basis set (AWCV5Z) leads to excellent agreement (1.81 eV). Subsequent test calculations (not shown here) on the smaller clusters ($n0$, $n1$, $n2$ and $n3a$) have revealed that additional diffuse functions have only limited effects (up to 5 meV) where MP2/AVQZ calculations still predict the $n3b$ isomer to be more stable than the $n3a$.

\FloatBarrier

\subsection{H+OH(H$_{2}$O)$_{n}^{-}$ collisional complex}

\begin{figure}[]
\centering
\begin{subfigure}{0.49\textwidth}
\includegraphics[width=1\textwidth]{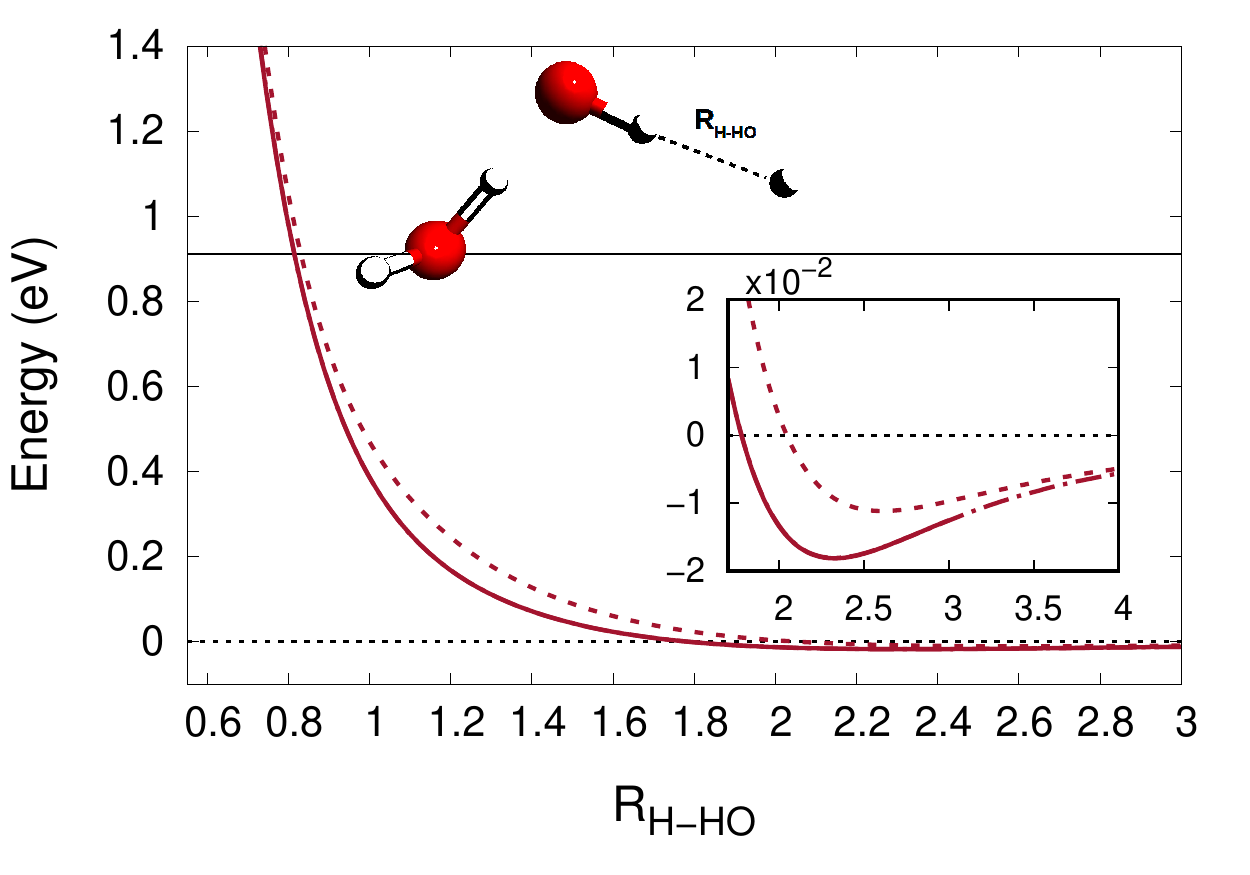}%
\end{subfigure}
\begin{subfigure}{0.49\textwidth}
\includegraphics[width=1\textwidth]{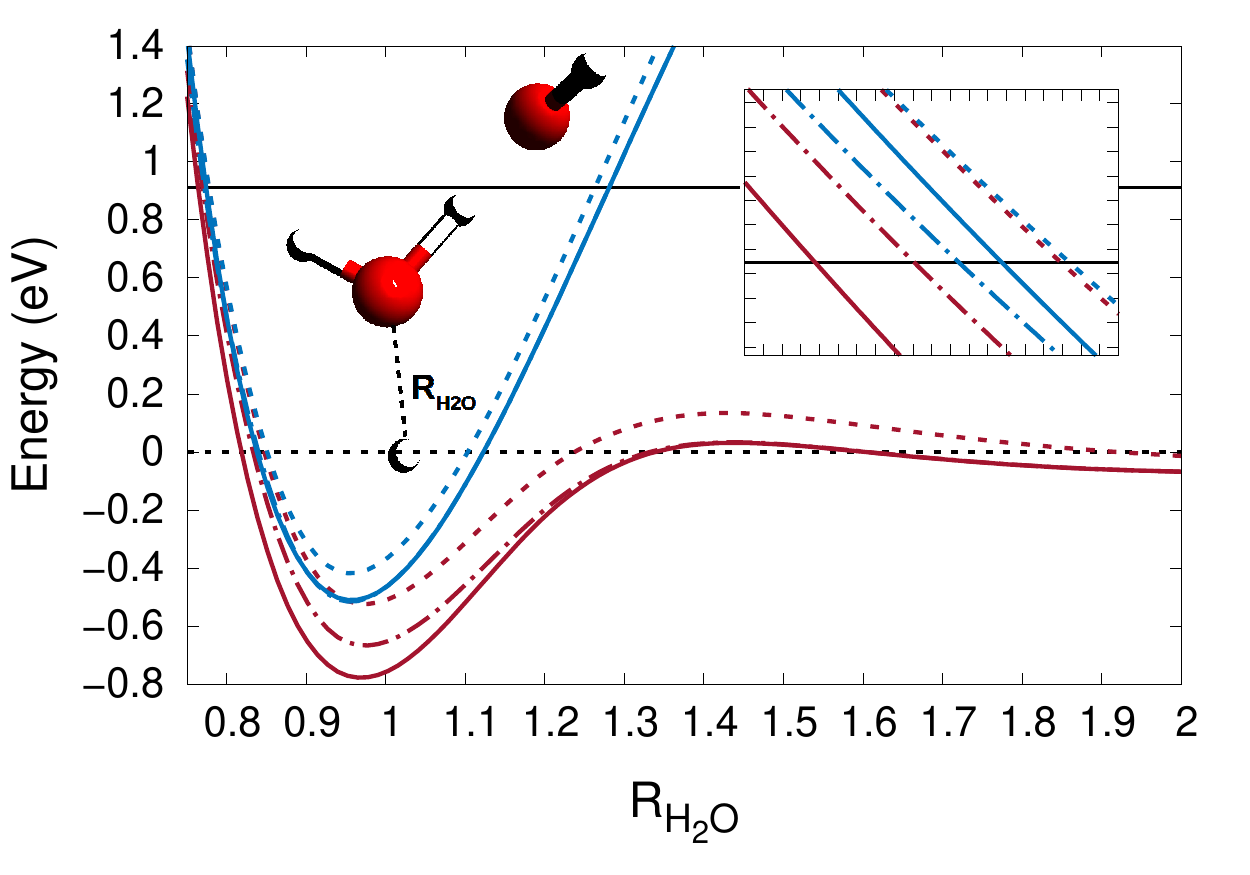}%
\end{subfigure} \\
\begin{subfigure}{1\textwidth}
\includegraphics[width=1\textwidth]{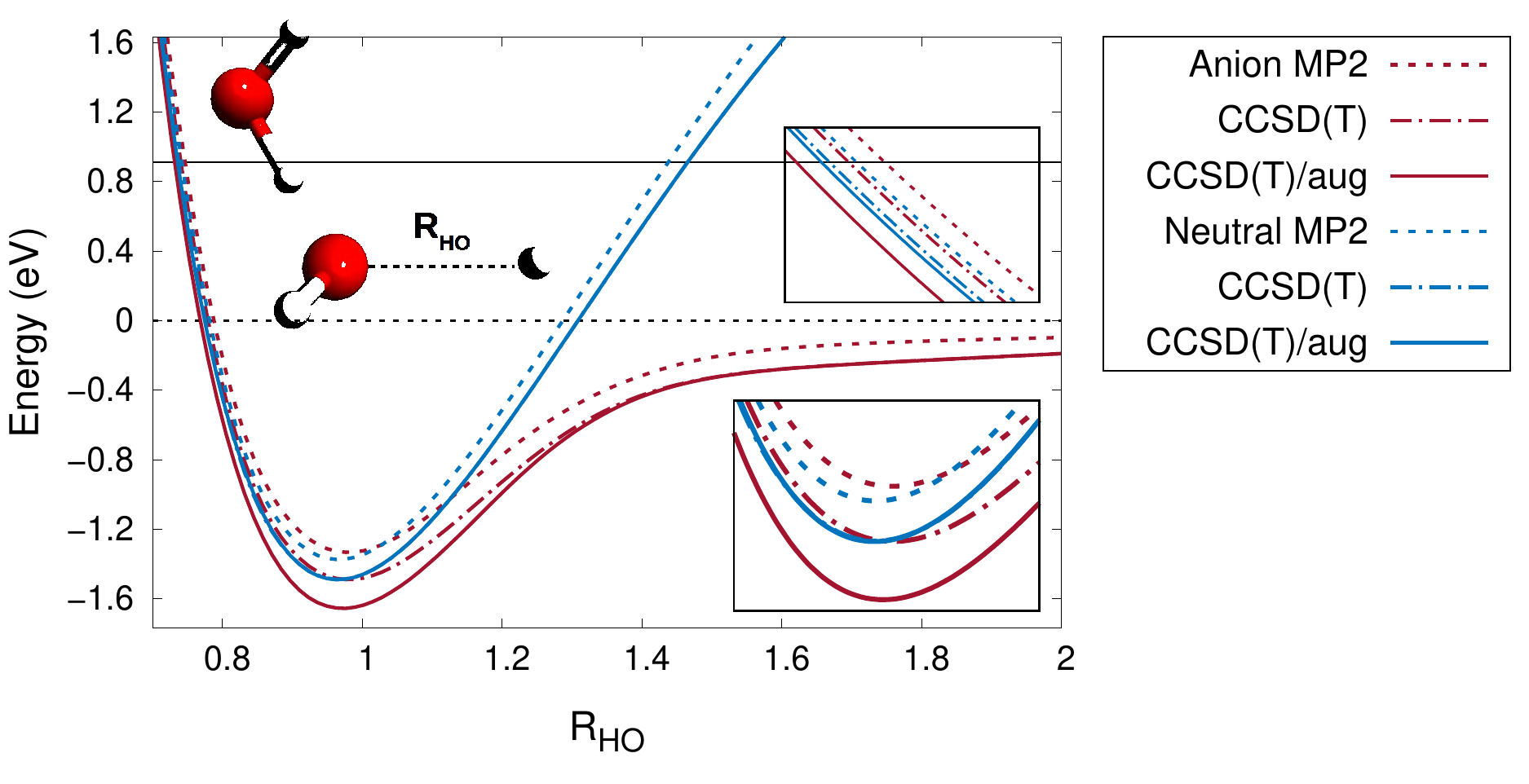}
\end{subfigure}
\caption{MP2 and CCSD(T) neutral (blue) and anion (red) PECs of the H-OH(H$_{2}$O)$^{-}$ molecular specie as a function of the H-HO distance (upper left), H-OH$_{2}$ distance (upper right) and H-OH distance (bottom). The coordinates are shown in the molecular cartoons. The internal coordinates of the cluster have been kept frozen at their MP2/AVTZ/AVTZ optimized value. The neutral PEC in the left panel is not seen as it lies above 4 eV. The insets show the small attractive potential at intermediate distances for H-HO and the behaviour of the PECs in the repulsive region for the two other cases. Also shown is the minimum for the H-OH case.}
\label{PEC_Hn1}
\end{figure}

\begin{figure}[]
\centering
\includegraphics[scale=0.8]{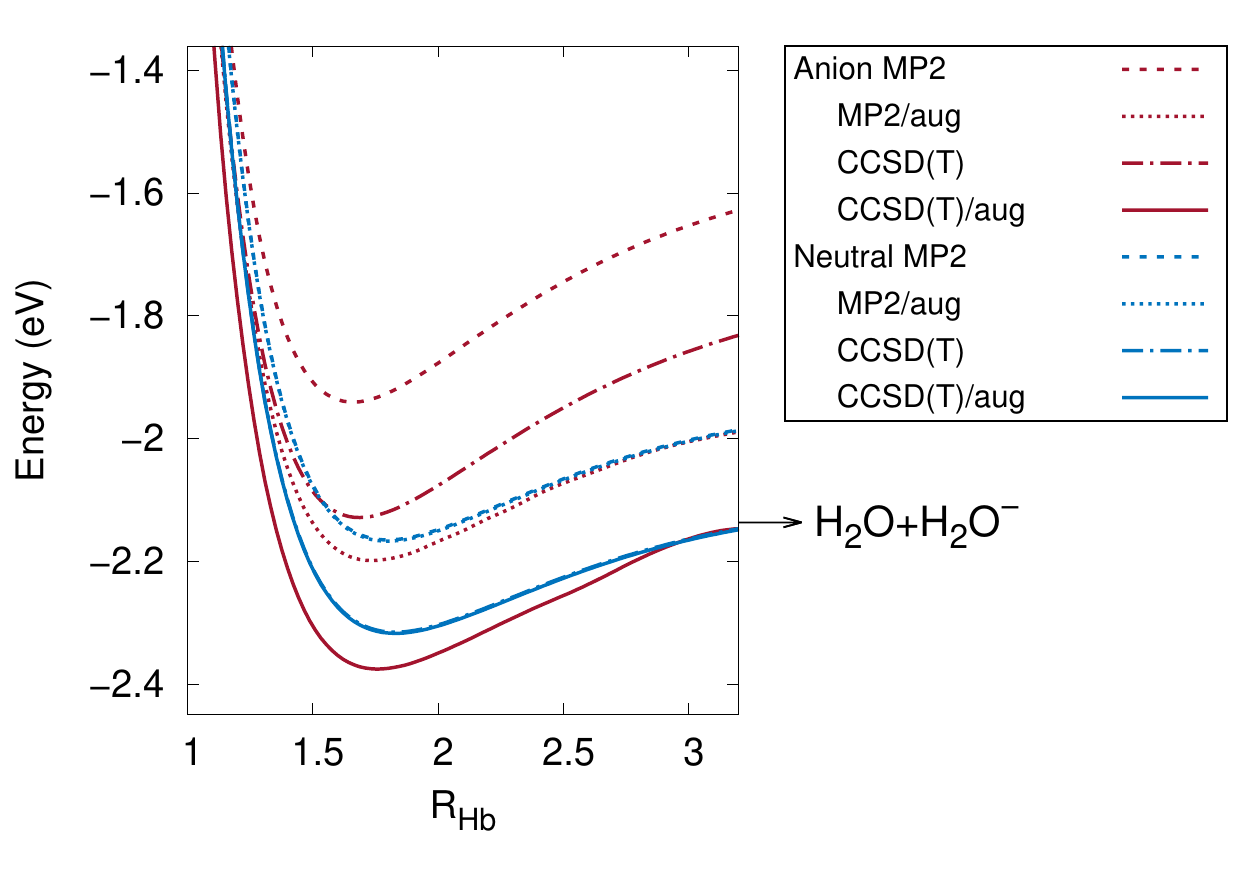}
\caption{MP2 and CCSD(T) PEC for H-OH(H$_{2}$O)$^{-}$ along the dissociation coordinates of the water group (R$_{Hb}$). Both AVTZ/AVTZ and AVTZ/AVTZ+aug results are shown.}
\label{Hn1_RHb}
\end{figure}

Figure \ref{PEC_Hn1} shows the PECs associated to 3 different collisional approaches of the H atom on the OH(H$_{2}$O)$^{-}$ target: from the H side of the OH group (upper right inset), from the O side of the H$_{2}$O group (upper left) and from the O side of the OH group (bottom). The later corresponding to the preferred approach. We have added $5s$, $3p$ and $1d$ even tempered functions on the incoming H atom. This choice is justified by the strong $1s_{H}$ nature of the HOMO when the H-OH(H$_{2}$O)$^{-}$ complex is formed. Although the atomic orbitals of O also contribute to the HOMO, we found the effect of adding diffuse functions on the O atoms to be negligible.
As already pointed out on the basis of the PES cuts in Figure \ref{Map_Hn1} (section \ref{sec_Hn1} in the main text), the configuration where the H atom approaches the O atom of the OH group (bottom in Figure \ref{PEC_Hn1} leads to largest interaction energy. This thus corresponds to the "preferred" collisional path. The collisional approaches targeting the water group or the H atom of the OH group gives rise to much higher lying autodetachment region. This observation supports the "shielding" effect of the water group hypothesis, resulting in a decrease of the AED reaction rate in comparison to the non hydrated case (H+OH$^{-}$). The use of more diffuse basis stabilises the anion which leads to higher lying AD region. The result seen in Figure \ref{Map_Hn1}, which were obtained without additional diffuse functions, thus corresponds to a lower limit of the AD region. It is interesting to note that these additional diffuse functions only slightly effect the PECs corresponding to the incoming H approaching the H atom from the hydroxide group (upper right panel in Figure \ref{PEC_Hn1}). This can be explained by the very different nature of the HOMO for the two different configurations. In the H-OH situation, the HOMO exhibits large contributions from diffuse $1s_{H}$ atomic orbitals whereas in the H-HO case, the HOMO is mainly formed by more contracted $2p_{O}$ atomic orbitals. The latter configuration is thus not a dipole-bound state, which is confirmed by a calculated dipole moment of around 1.85 D while the values for both other configurations are above 10 D. The BSSE effect on the interaction energy is around 0.1 eV, hence comparable to the method and basis set effects. However, it only slightly affects the VDE and AD region since the correction is almost the same for the anion and neutral case. \\
\noindent In order to investigate the indirect detachment process leading to the H$_{2}$O+H$_{2}$O$^{-}$ products, we have calculated the PEC as a function of the R$_{Hb}$ distance. The other coordinates have been kept frozen. The dissociation limit is the H$_{2}$O+H$_{2}$O$^{-}$ channel, where the water anion is unstable. Results can be seen in Figure \ref{Hn1_RHb}.
Once again, the critical effects of methods and diffuse functions can be seen. In particular, the H-OH(H$_{2}$O)$^{-}$ complex is predicted to be stable only when including the additional diffuse functions. The results for larger distance tend to the value calculated for H$_{2}$O$^{-}$ in table \ref{EAEint_Hcluster}. Both neutral and anion PEC crosses around R$_{Hb}=3 \angstrom$. \\

\FloatBarrier

\subsection{Isomer dependencies}

\begin{figure}
\begin{subfigure}{0.4\textwidth}
\includegraphics[scale=0.6]{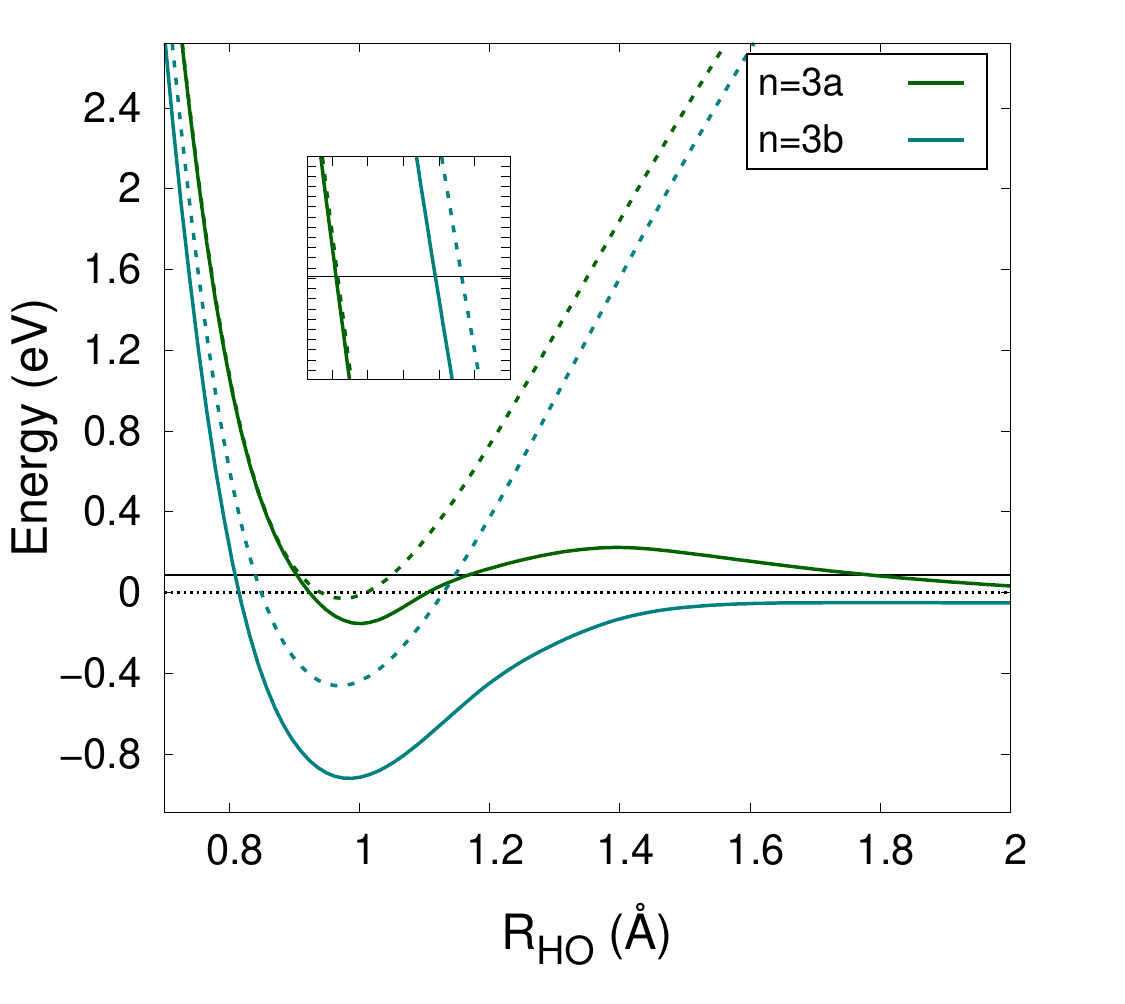}
\end{subfigure}
\hspace{30pt}
\begin{subfigure}{0.4\textwidth}
\includegraphics[scale=0.6]{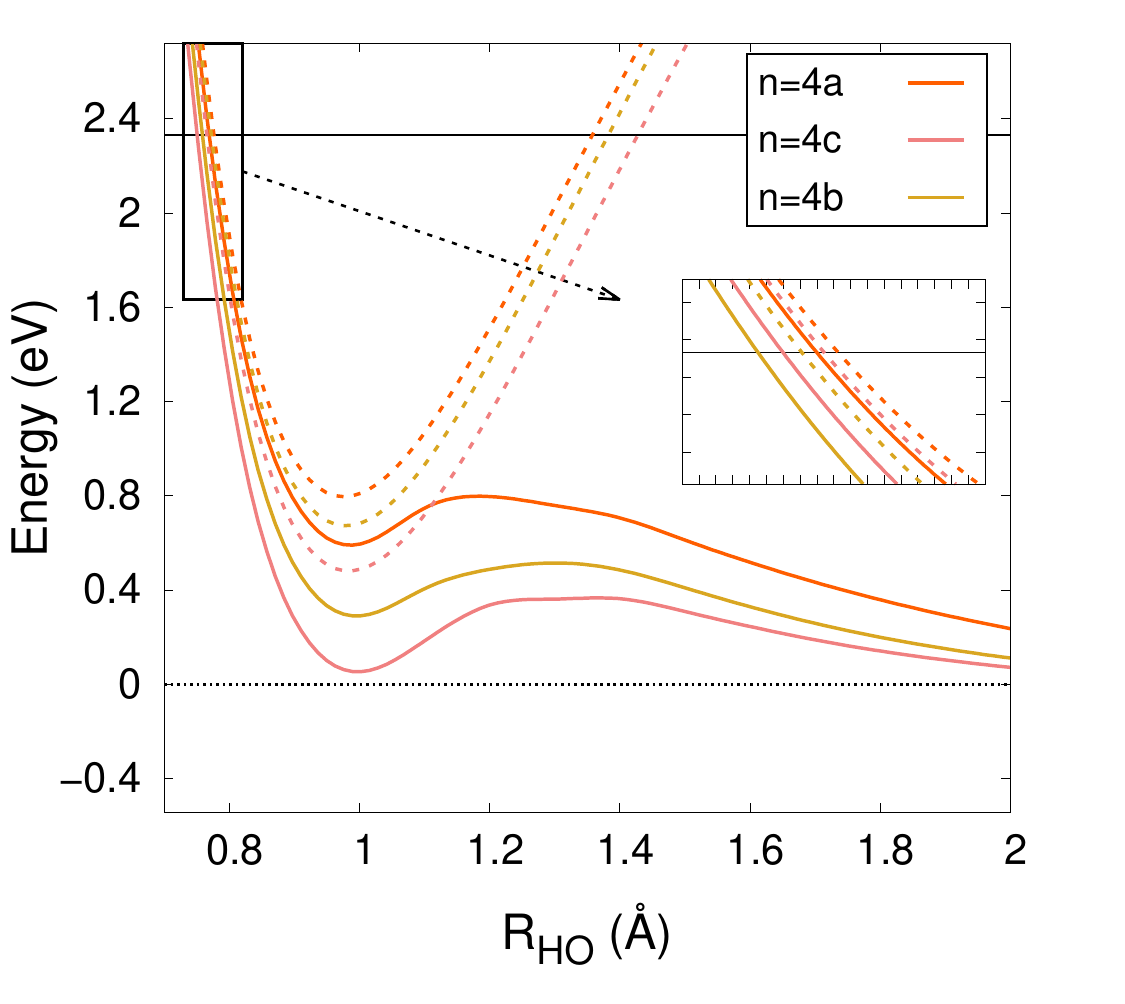}
\end{subfigure}
\caption{Preferred approach MP2/AVTZ/AVQZ+aug potential energy curves for the anion (solid lines) and neutral (dashed lines) H+OH(H$_{2}$O)$_{n}^{-}$ collisional system for $n=3a$, $3b$, $4a$, $4b$ and $4c$. The black dashed line corresponds to the energy at dissociation while the full black line represents the energy at the entrance channel (E$_{0}$, see main text for details).}
\label{Cross_H_iso}
\end{figure}

\begin{figure}
\begin{subfigure}{0.4\textwidth}
\includegraphics[scale=0.6]{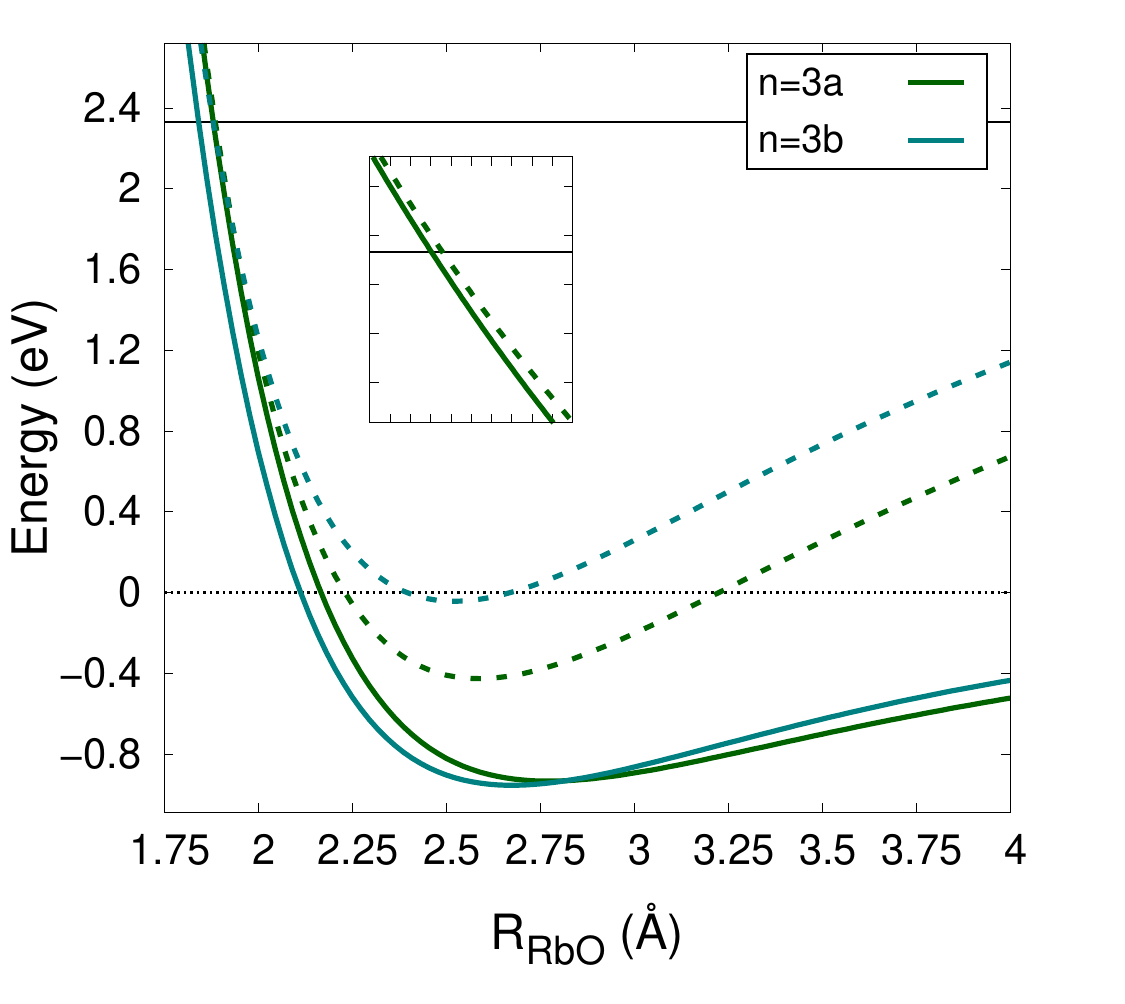}
\end{subfigure}
\hspace{30pt}
\begin{subfigure}{0.4\textwidth}
\includegraphics[scale=0.6]{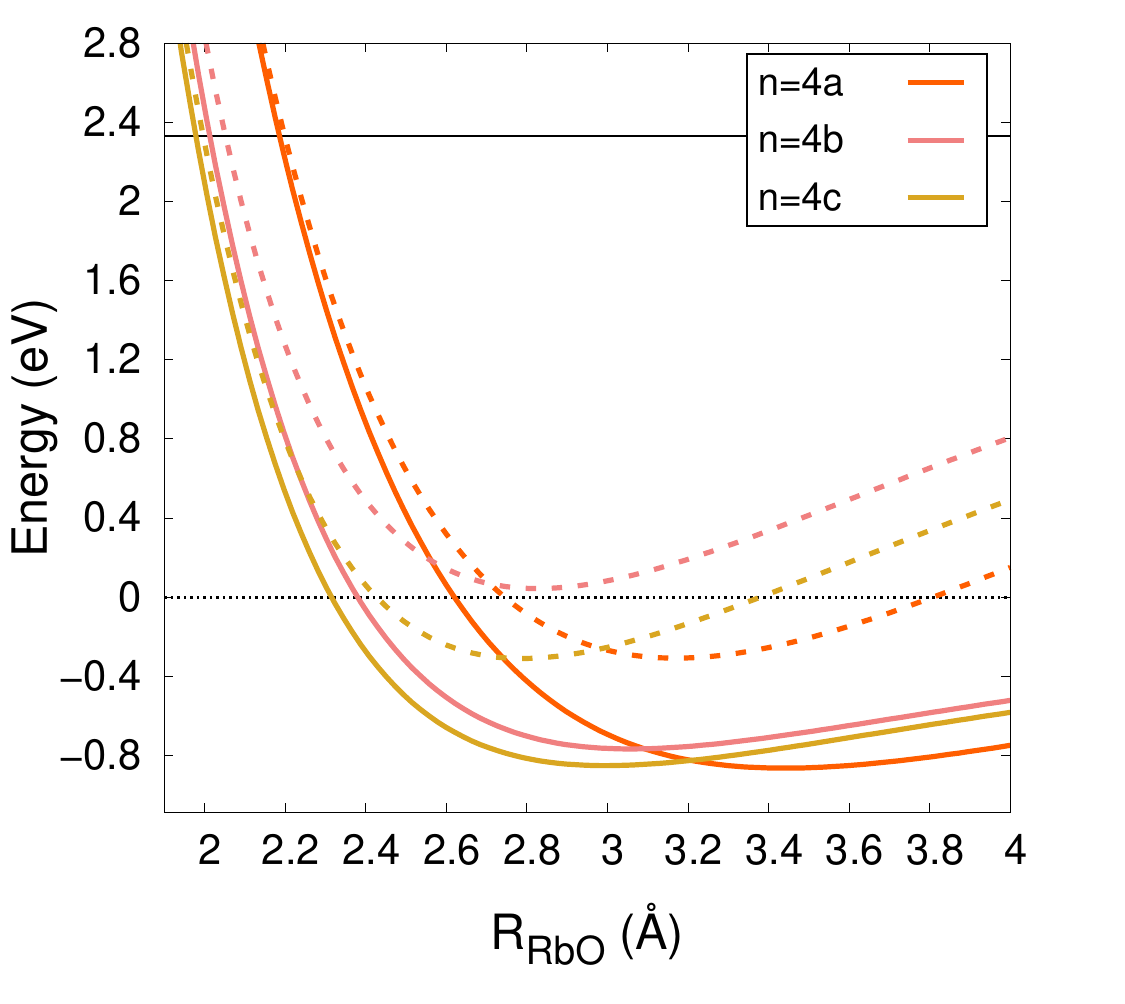}
\end{subfigure}
\caption{Preferred approach MP2/AVTZ/MDF$spdfg$+aug potential energy curves for the anion (solid lines) and neutral (dashed lines) Rb+OH(H$_{2}$O)$_{n}^{-}$ collisional system for $n=3a, 3b, 4a, 4b$ and $4$c. The black dashed line corresponds to the energy at dissociation while the full black line is the energy at the entrance channel (E$_{0}$, see main text for details).}
\label{Rbn_iso}
\end{figure}

The PECs corresponding to the preferred approach for the different isomers are shown in Figure \ref{Cross_H_iso} and \ref{Rbn_iso} for H and Rb, respectively .  

\end{document}